\documentclass[sn-mathphys,Numbered]{sn-jnl} 

\usepackage{graphicx}%
\usepackage{multirow}%
\usepackage{amsmath,amssymb,amsfonts}%
\usepackage{amsthm}%
\usepackage{mathrsfs}%
\usepackage[title]{appendix}%
\usepackage{xcolor}%
\usepackage{textcomp}%
\usepackage{manyfoot}%
\usepackage{booktabs}%
\usepackage{algorithm}%
\usepackage{algorithmicx}%
\usepackage{algpseudocode}%
\usepackage{listings}%
\usepackage{pdfpages}
\usepackage{lmodern}
\usepackage{hyperref}
\usepackage{lineno}
\begin{document}

\title[SPACIER]{SPACIER: On-Demand Polymer Design with Fully Automated All-Atom Classical Molecular Dynamics Integrated into Machine Learning Pipelines}

\author*[1]{\fnm{Shun} \sur{Nanjo}}\email{nanjos@ism.ac.jp}
\author[2]{Arifin}
\author[3]{\fnm{Hayato} \sur{Maeda}}
\author[1,4]{\fnm{Yoshihiro} \sur{Hayashi}}
\author[3]{\fnm{Kan} \sur{Hatakeyama-Sato}}
\author[1]{\fnm{Ryoji} \sur{Himeno}}
\author[3]{\fnm{Teruaki} \sur{Hayakawa}}
\author*[1,4]{\fnm{Ryo} \sur{Yoshida}}\email{yoshidar@ism.ac.jp}

\affil[1]{The Graduate University for Advanced Studies, SOKENDAI, Tachikawa, Tokyo, 190-8562, Japan}
\affil[2]{RD Technology and Digital Transformation Center, JSR Corporation, Kawasaki, 210-0821, Japan}
\affil[3]{Tokyo Institute of Technology, Meguro-ku, Tokyo 152-8550, Japan}
\affil[4]{The Institute of Statistical Mathematics, Research Organization of Information and Systems, Tachikawa, Tokyo 190-8562, Japan}

\abstract{
Machine learning has rapidly advanced the design and discovery of new materials with targeted applications in various systems. First-principles calculations and other computer experiments have been integrated into material design pipelines to address the lack of experimental data and the limitations of interpolative machine learning predictors. However, the enormous computational costs and technical challenges of automating computer experiments for polymeric materials have limited the availability of open-source automated polymer design 
systems that integrate molecular simulations and machine learning. We developed SPACIER, an open-source software program that integrates RadonPy, a Python library for fully automated polymer physical property calculations based on all-atom classical molecular dynamics into a Bayesian optimization-based polymer design system to overcome these challenges. As a proof-of-concept study, we successfully synthesized optical polymers that surpass the Pareto boundary formed by the tradeoff between the refractive index and Abbe number.
}

\maketitle

\section*{Introduction}\label{sec1}

Over the past decade, machine learning has shown significant potential for accelerating the discovery of new materials for numerous material systems. Machine learning algorithms for the on-demand design of new materials with desired properties have attracted considerable attention. Conventional machine learning pipelines comprise two steps for solving forward and inverse problems \citep{agrawal2016perspective}. In the forward problem, a machine-learning predictor is trained on a given dataset, defining the forward mapping from the composition and structural features of any given material to its properties. In contrast, in the inverse problem, the inverse mapping of the forward model is explored to backwardly predict materials exhibiting a given set of desired properties. This concept is general and applicable to a broad range of tasks in material research. Machine-learning pipelines have been successfully used to discover new materials across diverse material systems, including polymers \citep{wu2019machine}, inorganic compounds \citep{merchant2023scaling, szymanski2023autonomous}, alloys \citep{rao2022machine}, catalysts \citep{zhong2020accelerated, kim2019artificial}, and quasiperiodic materials \citep{liu2021machine, liu2023quasicrystals, uryu2024deep}.

A major challenge in data-driven materials research is the lack of data resources. In many cases, obtaining sufficient data for machine learning applications is difficult. Additionally, the ultimate goal of materials science is to discover ``innovative’’ materials from unexplored spaces with little or no available data. In particular, the scarcity of data on polymeric materials is remarkable. Currently, the most comprehensive polymer property database is PoLyInfo, which compiles around 100 properties of approximately 20,000 polymers from literature \citep{ishii2024nims}. However, applying PoLyInfo to machine learning is challenging because batch downloading via API is prohibited. Furthermore, only a few dozen entries are available for most properties, with the exception of a few basic properties, such as the glass transition and melting temperatures. For example, the number of samples required for the thermal conductivity near room temperature is fewer than 30 \citep{wu2019machine}. Other databases, such as the polymer property predictor and database \citep{kim2018polymer} and the polymerization reaction database CoPolDB \citep{takahashi2024copoldb}, also suffer from limited sample sizes.

Computer experiments have been integrated into machine-learning pipelines to overcome the quantitative limitations of experimental data and the hurdle of interpolative machine-learning predictions. Various machine learning algorithms have been developed for inorganic solid-state materials and small molecules that integrate \textit{ab initio} electronic structure calculations, such as the density functional theory. Experimental design methods, such as Bayesian optimization (BO) \citep{brochu2010tutorial, shahriari2015taking, frazier2018tutorial}, adaptively refine the machine-learning surrogate of physics-based simulation models, allowing efficient searches for materials with the desired properties while reducing the number of required computer experiments. Various examples of BO-aided computer experiments have been demonstrated, including enhancing heat transfer in bulk \citep{seko2015prediction} and nanostructured materials \citep{ju2017designing}, crystal structure prediction using first-principles calculations \citep{yamashita2021cryspy}, computational fluid dynamics of solids and fluids \citep{tran2019pbo}, composition optimization of wavelength-selective multilayer thermal radiation films \citep{sakurai2019ultranarrow}, and the design of fluorescent small molecule materials \citep{sumita2018hunting}.

However, the research on polymeric materials has been hindered owing to technical barriers in automating and accelerating all-atom molecular simulations. Therefore, previous studies have dealt with coarse-grained models, limiting the properties and polymer systems analyzed. Wang et al. (2020) \citep{wang2020toward} used BO-integrated coarse-grained molecular dynamics (MD) simulations to determine the particle sizes and intermolecular interaction strengths that enhance the ionic conductivity of solid polymer electrolytes and then back-mapped the estimated parameters to polymer species. Wu et al. (2023) \citep{wu2023coarse} applied BO to fit coarse-grained model parameters to experimental observations. 

Here, we developed an autonomous polymer design tool, materials SPAce frontier (SPACIER), which integrates fully automated polymer physical property calculations based on all-atom classical MD simulations into a BO-accelerated material design pipeline. RadonPy \citep{hayashi2022radonpy} is open-source software that can fully automate polymer physical property calculations using MD simulations. Given a polymer repeat unit, degree of polymerization, and other calculation conditions, the entire MD simulation process is fully automated, including conformational search, charge calculation, force field parameter assignment, polymer chain generation, equilibrium and nonequilibrium calculations, and physical property calculations. The main engine for the MD simulations was constructed using the Large-scale Atomic/Molecular Massively Parallel Simulator (LAMMPS) software. SPACIER implemented a set of codes to build an automated polymer design workflow using RadonPy. Using the various acquisition functions implemented in SPACIER, we can perform ordinary black-box and multi-objective optimizations or stochastic enumeration of polymers in any given property region.

As shown below, SPACIER can autonomously and comprehensively identify polymers that constitute the Pareto frontier or a desired region of experimental properties that can be calibrated from the property space computable with RadonPy. Furthermore, linking with sophisticated external molecule generators such as SMiPoly \citep{ohno2023smipoly}, a virtual library generator based on polymerization reaction rules, makes it possible to design highly synthesizable polymers while guiding their synthetic routes. In this paper, we present several examples of using SPACIER. In particular, we explored optical polymers that simultaneously enhanced the refractive index and Abbe number. The Abbe number is a physical property that describes the color dispersion of a transparent material, i.e., the change in the refractive index with wavelength. There is a tradeoff between these two properties, forming the Pareto frontier. As a proof-of-concept study, we used a multi-objective optimization algorithm of SPACIER to predict and successfully synthesize optical polymers exceeding the empirically known Pareto boundary of the refractive index and Abbe number.

\clearpage

\section*{Results}\label{sec2}
\subsection*{Methods outline}

\begin{figure}[h]%
\centering
\includegraphics[width=1\textwidth, page=1]{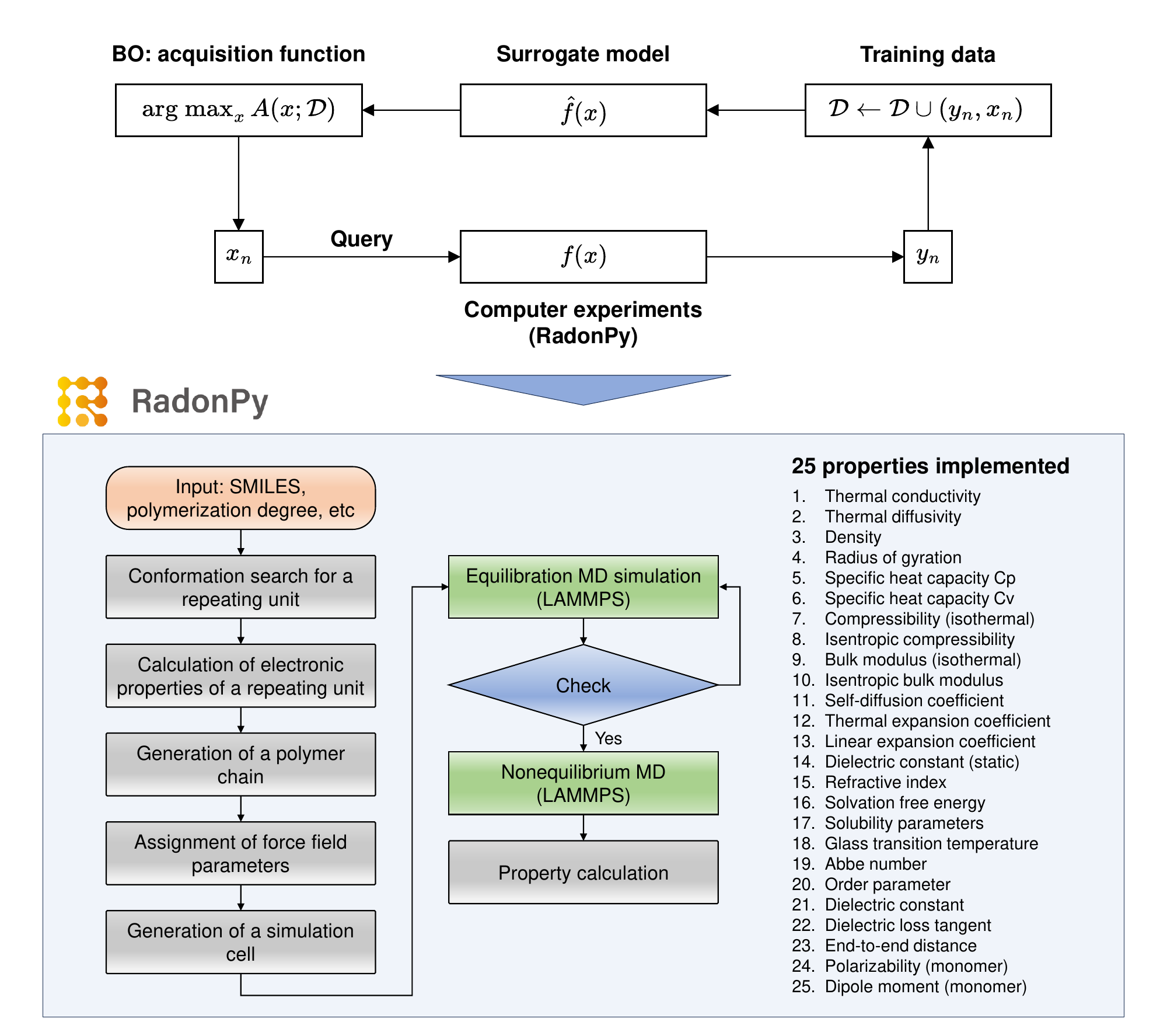}
\caption{SPACIER workflow: Bayesian optimization (BO) is utilized to identify polymers with desired properties. High-throughput calculation of polymeric properties is conducted using RadonPy, a fully automated tool for all-atom classical molecular dynamics (MD) simulations. The latest version of RadonPy implements automatic calculation algorithms for 25 different properties. This study considers the $C_\text{p}$, refractive index and Abbe number as the target properties.}
\label{fig:Fig1}
\end{figure}

We built a machine learning workflow incorporating BO with automated polymer physical property calculations using RadonPy (Fig. \ref{fig:Fig1}). With a given library of virtual polymers generated as described later, a pool-based BO was applied to identify promising candidates with the desired properties. For each polymer, the compositional and structural features of the repeating unit were translated into a 170-dimensional descriptor using a force-field kernel mean descriptor \citep{kusaba2023representation}. A Gaussian process (GP) surrogate $Y = f(X)$ with a Gaussian radial basis function kernel approximates the mapping from the vectorized polymer $X$ to the MD-calculated property $Y$. The candidate polymer that maximizes the calculated acquisition function was selected from the library, and its MD properties were then calculated. This input--output observation was added to the training dataset to retrain the surrogate model. This procedure was repeated until the polymers reaching the target properties were exhaustively explored.

In the two case studies presented below, RadonPy was used to evaluate three physical properties of amorphous polymers: the specific heat capacity at constant pressure ($C_\text{p}$), refractive index, and Abbe number. Hayashi et al. released the first version of RadonPy, which implemented automatic calculation algorithms for 14 properties, including $C_\text{p}$ and the refractive index. RadonPy is currently being developed as part of a consortium-based open-source project. In this study, we released an updated version that implements automatic calculation of the Abbe number (RadonPy version 0.2.3). In RadonPy, standardized calculation conditions, known as presets, have been implemented for various properties and polymer systems, as determined by experts based on the experimental properties. 
However, the MD-calculated properties did not match the experimental values perfectly. For example, $C_\text{p}$, as calculated by classical MD, was overestimated compared to the experimental values because of the absence of quantum effects. A linear calibrator was derived from the experimental and calculated data to collect the systematic bias.

The candidate polymer sets for the two applications consisted of 1,077 synthetic polymers provided by Hayashi et al. (2022) \citep{hayashi2022radonpy} and 101,487 virtual polymers generated using the rule-based polymerization reaction model SMiPoly. SMiPoly is a virtual polymer generator that implements 22 polymerization reaction rules, consisting of six-chain polymerization reactions and 16 step-growth polymerization reactions. Specifically, using 1,083 readily available monomers, 169,347 unique polymers were generated, forming seven different polymer types: polyolefin, polyester, polyether, polyamide, polyimide, polyurethane, and polyoxazolidone.

SPACIER implements the probability of improvement (PI) and expected improvement as acquisition functions for ordinary single-objective optimization. In the two examples presented, we performed multi-objective BO. In the first example, the following multi-objective version of the PI was used as an acquisition function to search for polymers reaching the desired property region:
\begin{eqnarray}
A(X, \mathcal{D}) = \prod_{k=1}^p \int_{l_k}^{u_k} p(Y_k | X, \mathcal{D}) \mathrm{d} Y_k.
\label{eq:eq1}
\end{eqnarray}
The acquisition function represents the probability that the $p$ target properties $(Y_1, \ldots, Y_p)$ belong to region $[l_1, u_1] \times [l_2, u_2] \times \ldots, [l_p, u_p]$ for the GP posterior predictive distribution $ p(Y_k | X, \mathcal{D})$. In another example, we searched for solution sets that lie on the Pareto boundary formed by the tradeoff between the refractive index and Abbe number. In SPACIER, expected hyper-volume improvement (EHVI) \citep{yang2019multi} is implemented as an acquisition function for multi-objective BO. 

The software interface of SPACIER operates as follows. The user sets an initial property dataset, candidate polymers, acquisition function type, and number of candidate polymers ($N$) to be selected in each BO step. SPACIER calculates the acquisition function based on the learned surrogate model and selects the top $N$ candidate polymers. Next, a job script to run RadonPy is automatically created, and the job is  submitted through the queuing system to obtain the MD-calculated properties. The surrogate is then re-learned using the newly added data. For further details, refer to the guidelines on the GitHub website \url{https://github.com/s-nanjo/Spacier}.

\subsection*{Illustrative example}
Herein, we describe the basic concept and utilization of SPACIER through its application to a simple toy problem. The target properties are $C_\text{p}$ and the refractive index. As depicted in the top panel of Fig. \ref{fig:Fig2}, the calculated values for $C_\text{p}$ overestimate the experimental values because of the absence of quantum effects in classical MD simulations. The refractive index calculated using RadonPy slightly underestimates the experimental values, likely owing to an underestimation of the density. The linear models were fitted to the experimental values to correct for these systematic biases (Fig. \ref{fig:Fig2}, bottom). The mean absolute errors were 167.53 and 0.02, and the coefficients of determination were 0.61 and 0.92 for the  $C_\text{p}$ and refractive index, respectively.

We used 1,077 polymers provided in the original RadonPy paper \citep{hayashi2022radonpy} as the candidate polymer set. Their MD properties were calculated to define the ground truth set for performance evaluation. As shown in Fig. \ref{fig:Fig2}a, the three target property ranges were located near the Pareto boundary of the joint distribution of the two properties for the 1,077 polymers. SPACIER was used to exhaustively identify the polymers.

We compared three machine learning methods: BO, Fix-GP, and Random. BO calculates the probability that the properties of each candidate polymer fall into the target region using the posterior predictive distribution of GP according to Equation \ref{eq:eq1}. In each step, the top 10 polymers in the acquisition function were selected, and the GP was sequentially retrained using additional property data. In the initialization step, GP was trained using 10 randomly selected polymers and their properties (Fig. \ref{fig:Fig3}a). Fix-GP performed polymer selection using the acquisition function, without updating the model trained on the initial dataset. In the Random method, 10 polymers were randomly selected at each step, serving as a control experiment.

BO detected all polymers in the three target regions within 20--30 cycles (Fig. \ref{fig:Fig3}b), demonstrating a clear advantage over Fix-GP and Random. The selected polymers were smoothly distributed to encompass the neighborhood of the target region (kernel density estimation in Fig. \ref{fig:Fig3}c). In general, there is a discrepancy between computational models and a real-world system; therefore, the optimal solution in the computer experiment does not coincide with that in a real system. Therefore, it is vital to enumerate the optimal solution and its search path during hill climbing as well as the neighborhood distribution exhaustively and unbiasedly, facilitating unbiased decision-making by experts.

Examples of identified polymers in each region are shown in Fig. \ref{fig:Fig3}d. Region 1 ($[1000, 1500] \times [1.75, 1.85]$ for $C_\text{p}$ and refractive index, respectively), which has a relatively low $C_\text{p}$ and high refractive index, contains many conjugated polymers rich in aromatic rings. 
In Region 2 ($[1500, 2000] \times [1.60, 1.70]$), numerous aromatic polymers with sp$^2$ carbons as building blocks were detected. Region 3 ($[2000, 2500] \times [1.50, 1.60]$) predominantly features polymers rich in sp$^3$  carbons. Thus, SPACIER can comprehensively search for polymers with desired physical properties using RadonPy, even in the absence of experimental data.

We also performed several ablation studies. When increasing the initial dataset size to 100, the detection performance of Fix-GP approached that of BO (Fig. S2). However, when the initial dataset was sampled from a biased region with low $C_\text{p}$ and a refractive index, the performance of Fix-GP was significantly lower, as expected (Fig. S3). Even when the target properties were changed, BO's performance remained significantly better than that of the baselines (Fig. S4). In addition, experiments using EHVI to search for the optimal solution set on the Pareto boundary of the two properties showed that BO could detect all solutions in approximately 30 cycles (Fig. S5).

\subsection*{Optical polymers predicted and discovered by SPACIER}

SPACIER was used to predict and synthesize polymers that exhibit high refractive index and Abbe number required for optical materials.
For example, allyl diglycol carbonate and polymethyl methacrylate, known for their high Abbe number and excellent processability, have been widely employed in eyeglass lenses. However, there is the empirical ``limiting boundary'' between the refractive index and Abbe number, formed by their tradeoff relationship \citep{okutsu2008sulfur, cai2015sulfonyl}. This study aimed to discover polymers going beyond the empirical limits of these two properties. 

\begin{figure}[h]%
\centering
\includegraphics[width=1\textwidth, page=1]{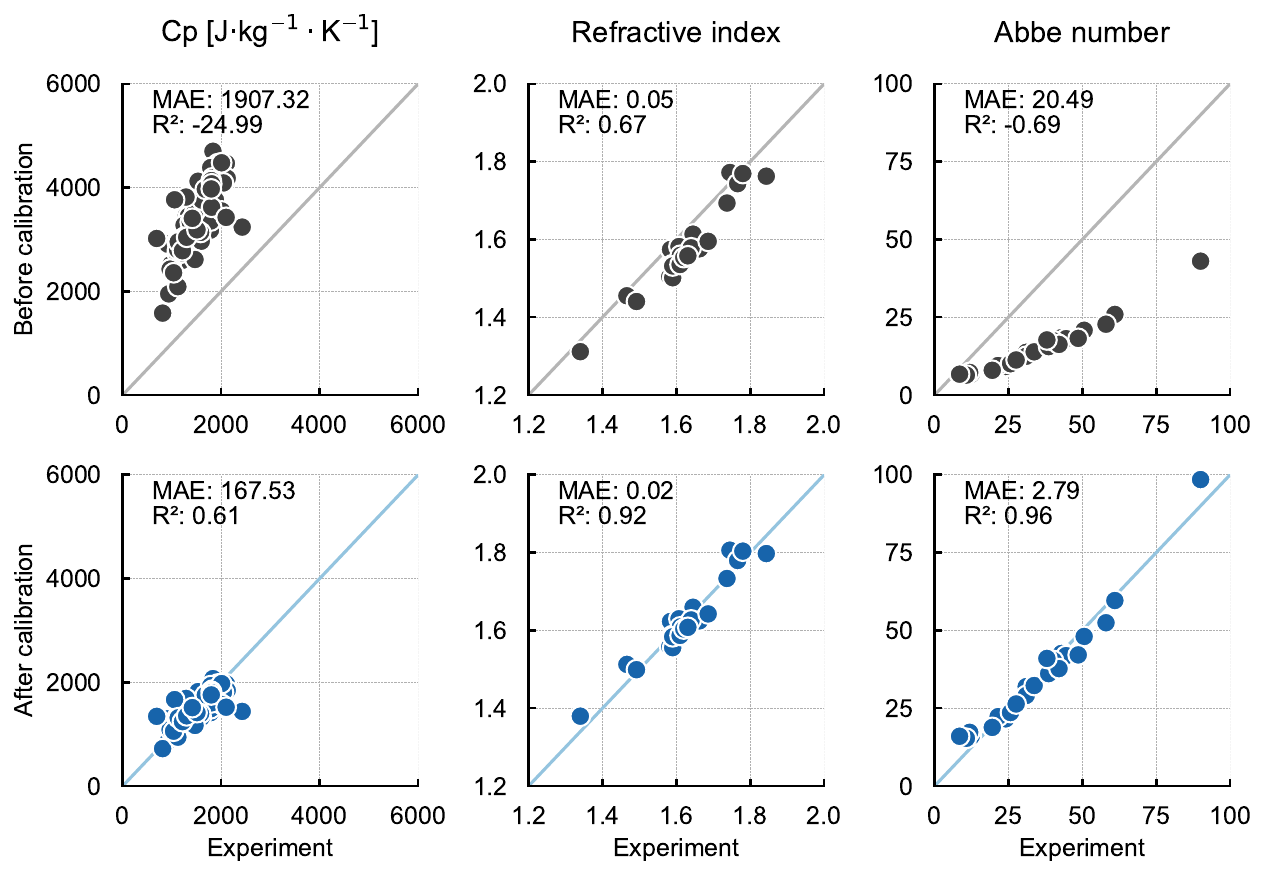}
\caption{Comparison of MD-calculated and experimental values for three physical properties ($C_\text{p}$, refractive index, and Abbe number). The horizontal axes represent the experimental values, while the vertical axes represent the MD-calculated properties (top) and the calibrated values from the linear models fitted to the experimental data (bottom).}
\label{fig:Fig2}
\end{figure}

\begin{figure}[h]%
\centering
\includegraphics[width=1\textwidth, page=1]{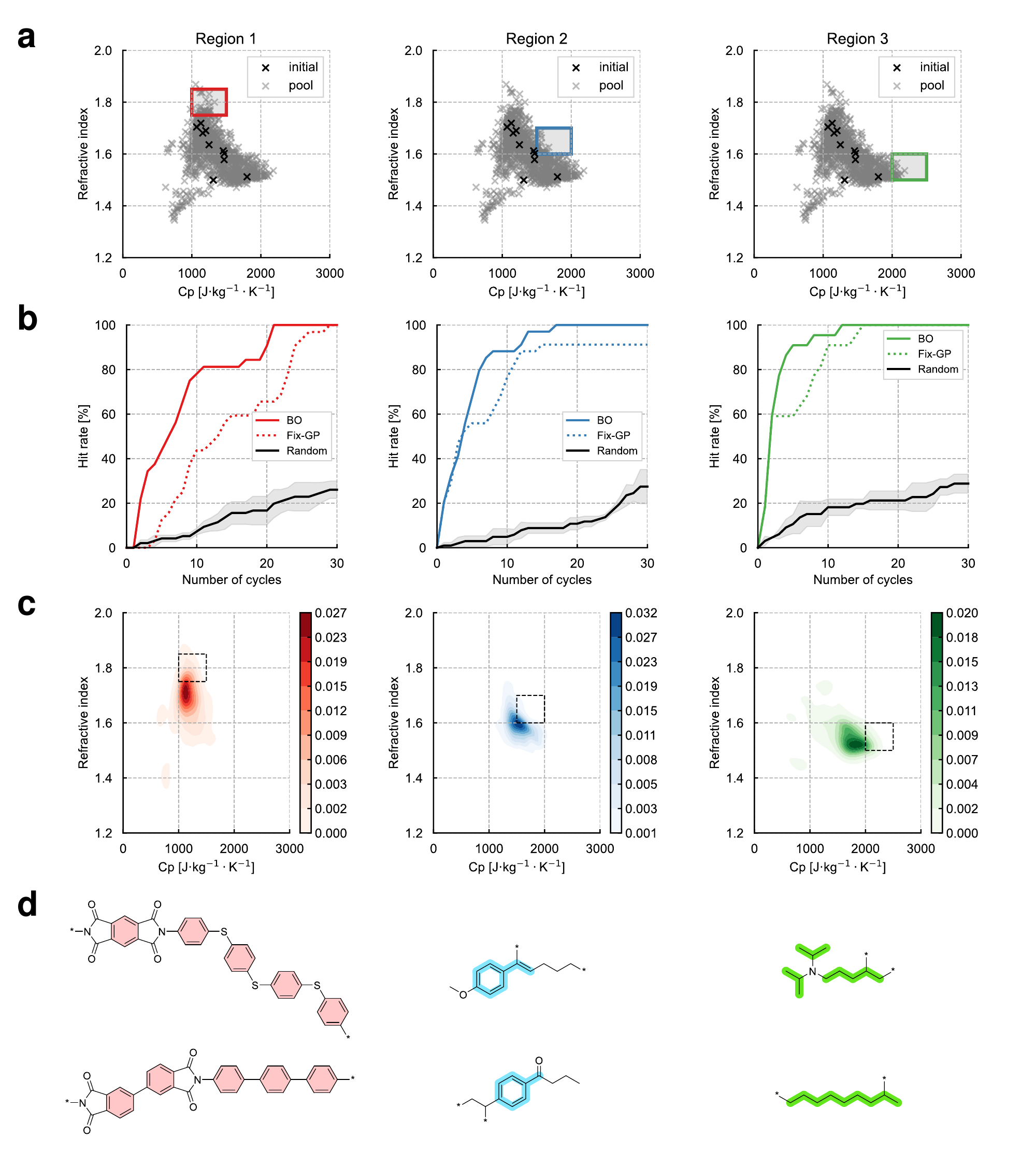}
\caption{
Example of SPACIER application targeting $C_\text{p}$ and refractive index. \textbf{a} Three different target property regions (enclosed by squares) are plotted on the joint distribution of the two MD-calculated properties for all candidate polymers (gray). Initial data points are plotted in black. \textbf{b} Hit rate versus the number of BO cycles. The hit rate represents the percentage of polymers within the designated target region. ``Random'' represents the mean and standard deviation of three independent trials. \textbf{c} Kernel density estimation of the MD-calculated properties for the polymers selected through BO. \textbf{d} Examples of polymers in each target region.
}
\label{fig:Fig3}
\end{figure}

Using SPACIER, we conducted a multi-objective BO with EHVI as the acquisition function. In each BO step, the top 10 polymers of the acquisition function were selected from the candidate polymers, which were polymerized using SMiPoly from 1,021 purchasable compounds. As depicted in the top panel of Fig. \ref{fig:Fig2}, the MD-calculated refractive indices and Abbe numbers underestimate the experimental values. Therefore, linear models were used to calibrate the MD properties (Fig. \ref{fig:Fig2}, bottom). The mean absolute errors were 0.02 and 2.79, and the coefficients of determination were 0.92 and 0.96 for the refractive index and Abbe number, respectively.

\begin{figure}[h]%
\centering
\includegraphics[width=1\textwidth, page=1]{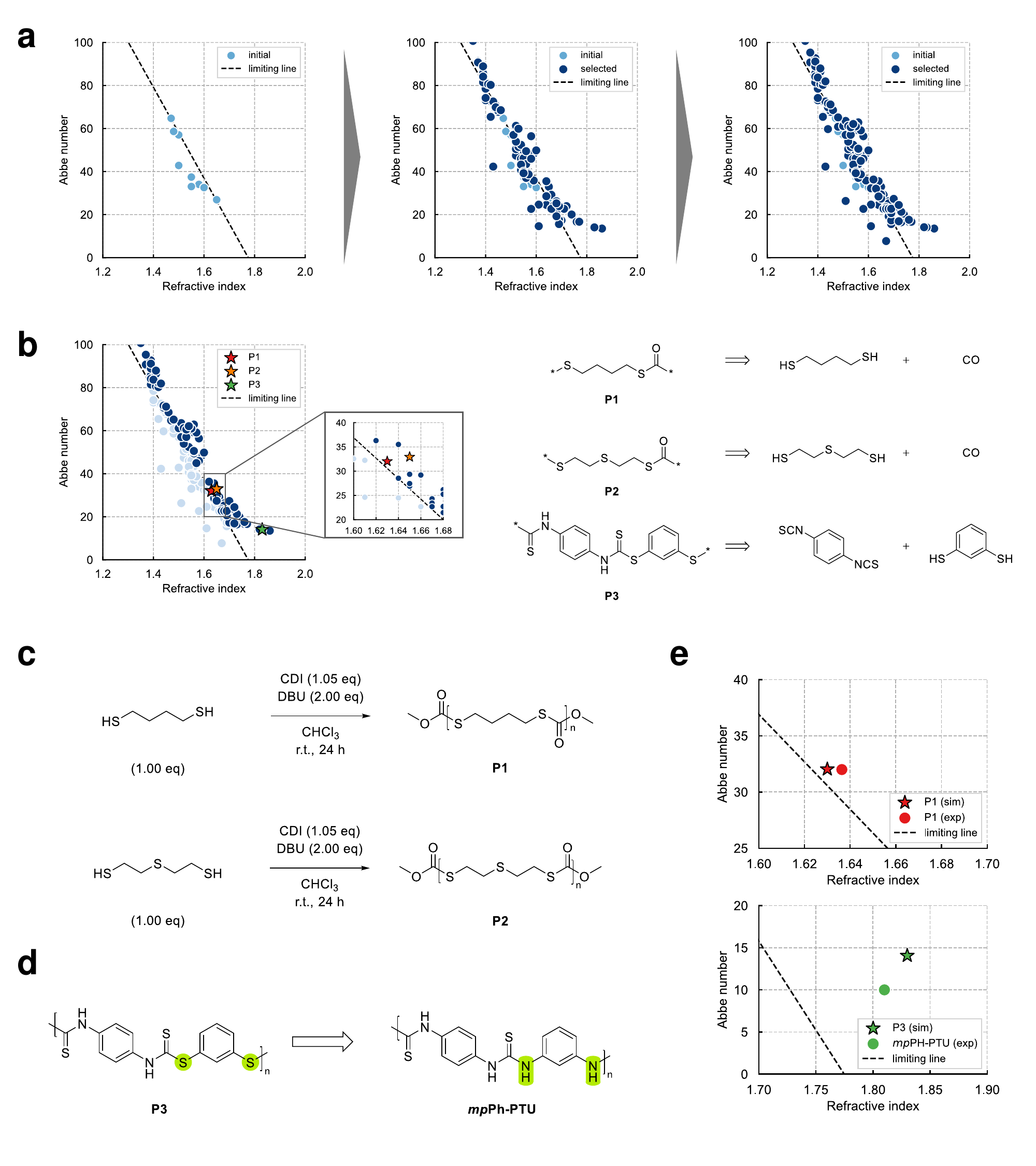}
\caption{
Prediction and synthesis of optical polymers. \textbf{a} MD-calculated properties of polymers accumulated through iterations of BO (left: initial distribution, center: step 10, right: step 20). The empirical limit boundary is indicated by a dashed line. This limit boundary was created based on Figure 7 from the literature \citep{okutsu2008sulfur}.
\textbf{b} Three polymers (\textbf{P1}, \textbf{P2}, \textbf{P3}) selected as synthetic targets, along with their polymerization reactions presented by SMiPoly. \textbf{c} Polymerization reactions and reaction conditions for \textbf{P1} and \textbf{P2}. \textbf{d} Structure of \textbf{\textit{mp}Ph-PTU}, selected as an analog of \textbf{P3}. \textbf{e} MD-calculated properties and experimental values of the newly synthesized \textbf{P1} and \textbf{\textit{mp}Ph-PTU}.
}
\label{fig:Fig4}
\end{figure}

During 20 cycles of the multi-objective BO, the designed polymers gradually approached and eventually crossed the empirically known Pareto frontier (Fig. \ref{fig:Fig4}a). The percentage of all polymers crossing the empirical limit line is 64 \%. Approximately one quarter of these polymers contain sulfur atoms with several sulfonyl groups (–SO$_2$–) as substructures. In previous studies \citep{cai2015sulfonyl, suzuki2012synthesis}, including sulfonyl groups into molecules was reported as a promising strategy for designing polymers that exceed the empirical limit. SPACIER has successfully learned this design principle autonomously.

For synthetic targets that go beyond the empirical boundary, we selected (poly)dithiocarbonate (\textbf{P1}, \textbf{P2}) and (poly)dithiourethane (\textbf{P3}) because the raw materials for these polymers were readily available (Fig. \ref{fig:Fig4}b). Of these three polymers, only the synthesis of \textbf{P1} has been previously reported \citep{berti1988sulfur}; however, its refractive index and Abbe number have not been reported. 
According to  SMiPoly's guide, \textbf{P1} and \textbf{P2} can be synthesized using a combination of dithiols and a carbonyl source (Fig. \ref{fig:Fig4}b). Common carbonyl sources include phosgene and diphenyl carbonate (DPC); however, phosgene is highly toxic and DPC requires harsh reaction conditions \citep{wnuczek2021synthesis}. Therefore, in this study, 1,1'-carbonyldiimidazole (CDI) was employed (Fig. \ref{fig:Fig4}c) following the method described in the literature \citep{sehn2022straightforward}. The detailed synthetic procedure is described in the Methods section.

During \textbf{P1} synthesis, the viscosity of the reaction mixture increased over time, forming the desired high-molecular-weight polymer. The structure was identified using nuclear magnetic resonance (NMR) and thin films were successfully fabricated on Si substrates using a spin-coating method.

During \textbf{P2} synthesis, a solid was precipitated during the polymerization reaction. \textbf{P2} was insoluble in commonly used solvents, preventing structural determination using NMR. To address this issue, we copolymerized the raw material of \textbf{P2}  with another monomer under the same reaction conditions (Fig. S13), improving product solubility (Table S1). However, the copolymer did not successfully form a film.

Subsequently, we attempted to synthesized \textbf{P3}. According to SMiPoly's guide, \textbf{P3} can be synthesized using a combination of diisothiocyanate and dithiol (Fig. \ref{fig:Fig4}b). However, based on reports of similar reactions \citep{yoshida2018synthesis}, the synthesis of the desired polymer could be difficult because of the low nucleophilicity of the aromatic dithiol. Fortunately, a recent study has reported the refractive index and Abbe number of \textbf{\textit{mp}Ph-PTU} \citep{watanabe2024polarizable}, a structural analog of \textbf{P3}. Therefore, instead of measuring the physical properties of \textbf{P3}, we referred to the reported physical property values of \textbf{\textit{mp}Ph-PTU} (Fig. \ref{fig:Fig4}d).

Table \ref{tab:Table1}  summarizes the experimental and MD-calculated properties. The values of the refractive indices and Abbe numbers for \textbf{P1} and \textbf{\textit{mp}Ph-PTU} are in good agreement. For the refractive index, the experimental values were 1.64 for \textbf{P1} and 1.81 for \textbf{\textit{mp}Ph-PTU}, while the MD-calculated values were 1.63 and 1.84, respectively. For the Abbe number, the experimental values were 32.0 for \textbf{P1} and 11.0 for \textbf{\textit{mp}Ph-PTU}, while the MD-calculated values were 32.0 and 14.1, respectively. The refractive indices and Abbe numbers of the structurally similar \textbf{P3} and \textbf{\textit{mp}Ph-PTU} also showed fairly close values for the experimental and calculated physical properties. Consequently, \textbf{P1} and the analogs of \textbf{P3} discovered by SPACIER exceed the currently known Pareto boundary in real-world systems (Fig. \ref{fig:Fig4}e).

\begin{table}[h]
\caption{Experimental and MD-calculated refractive indices and Abbe numbers for the three polymers predicted by SPACIER.}
\begin{tabular*}{\textwidth}{@{\extracolsep\fill}lcccc}
\toprule%
& \multicolumn{2}{@{}c@{}}{Refractive index} & \multicolumn{2}{@{}c@{}}{Abbe number} \\\cmidrule{2-3}\cmidrule{4-5}%
Polymer &  Experiment\footnotemark[1]\footnotemark[2] & Simulation  & Experiment\footnotemark[2] & Simulation \\
\midrule
\textbf{P1} & 1.64  & 1.63   & 32.0 & 32.0\\
\textbf{P3} & -  & 1.83   & - & 14.1\\
\textbf{\textit{mp}Ph-PTU}   & 1.81\footnotemark[3]  & 1.84   & 11.0\footnotemark[3] & 14.1\\
\botrule
\label{tab:Table1}
\end{tabular*}
\footnotetext[1]{Measured at 589 nm.}
\footnotetext[2]{Determined by spectroscopic ellipsometry.}
\footnotetext[3]{Literature values \cite{watanabe2024polarizable}.}
\end{table}

\clearpage

\section*{Discussion}\label{sec3}
We present the first proof-of-concept study of polymer design and synthesis using a machine-learning system incorporating automatic polymer physical property calculations based on all-atom MD simulations. As demonstrated through the two experiments, SPACIER is likely capable of reaching any region in the chemical space as long as the target polymer systems are computable in or calibratable from RadonPy. RadonPy has been undergoing expansion through a consortium-based open-source development. The potential of SPACIER can be further increased by extending RadonPy's functionality. 

Additionally, the synthesis process was accelerated using SMiPoly, a virtual library generator, with an exhaustive implementation of the polymerization reaction rules. Consequently, we successfully discovered two polymers that surpassed the Pareto boundary between the refractive index and Abbe number. The experimental and calculated properties of both polymers were aligned with sufficient accuracy. However, one polymer \textbf{P2}, although likely synthesized, failed to form a film due to its insolubility in organic solvents. In \textbf{P2}, a carbonyl group is partially inserted into poly(ethylene sulfide) (\textbf{PES}). \textbf{PES} is a known polymer that is insoluble in most organic solvents at room temperature \citep{oyama1999polymer}. Similarly to \textbf{PES}, \textbf{P2} has low solubility in organic solvents.

This study also highlights bottlenecks in the practical use of SPACIER and possible solutions. Only a few optical polymers have been found to significantly exceed the empirical boundary. This limitation was primarily due to the lack of structural diversity in the candidate polymers because we restricted our investigation to polymers that could, in principle, be synthesized in one step from commercially available monomers. For example, Ueda and Ando reported synthesizing polymers having high refractive indices (1.61--1.62) and Abbe numbers (48.0--45.8) in three or four steps including the monomer synthesis \citep{suzuki2012synthesis}. Applying SMiPoly with more compounds that can be synthesized in two or more reaction steps can add more diverse structures to the candidate set rather than limiting them to commercially available monomers. Another challenge is narrowing down the polymers that could be synthesized. For instance, the product obtained from the synthesis of \textbf{P2} was not soluble in the solvent. Excluding poorly soluble polymers in advance would allow the construction of a high-quality virtual library. The use of a machine learning solubility predictor can help address this issue. For example, Aoki et al. \citep{aoki2023multitask} demonstrated that a machine learning predictor could accurately predict the Flory--Huggins $\chi$ parameter of a polymer--solvent solution and determine whether an arbitrary polymer--solvent pair is soluble or insoluble. Considering these issues, we plan to update the software in future.

\section*{Methods}\label{sec4}
\subsection*{Candidate polymers}

As an illustrative example of SPACIER targeting $C_\text{p}$ and the refractive index, 1,077 polymers obtained from RadonPy's GitHub repository were used as the candidate polymer set.
To explore optical polymers, 101,487 virtual polymers were generated using the following procedure: 
\begin{enumerate}
\item[(1)] Readily available monomers were obtained from SMiPoly's GitHub repository.
\item[(2)] After removing cases involving cation-anion pairs, B atoms, or Si atoms, the extracted monomers were passed to SMiPoly for in silico polymerization reactions.
\item[(3)] Remove cases where polymers do not have two asterisks in the simplified molecular input line entry system (SMILES) string \citep{weininger1988smiles} from the generated polymers.
\item[(4)] Remove redundant polymers with identical repeating units using the ``poly.full\_match\_smiles\_listsel'' function of RadonPy.
\item[(5)] Remove cases where the number of atoms in the repeating unit is at least 55.
\item[(6)] Remove cases where the repeating unit matched those in the 1,077 polymers.
\end{enumerate}

\subsection*{Bayesian optimization}

The objective of BO is the derivative-free optimization of black-box function $f$ that maps the input $X$ of the system to the output $Y$. The optimal solution of $f$ is identified by sequentially generating the realizations of $X$ and $Y$, which improves the accuracy of the surrogate model as an estimator of $f$ while minimizing the total number of experiments. 

The BO procedure is summarized in Algorithm 1. It begins with an initial dataset $\{ (X_i, Y_i) | i =1, \ldots, n\}$, along with an acquisition function that aids decision-making for subsequent computer experiments. The key process involves selecting a query $X_{new}$ from a set of candidate polymers guided by the acquisition function to maximize potential improvements. Subsequently, a computer experiment observes the output $(Y_{new})$. This instance is added to the training dataset, and subsequently used to refine the surrogate model’s performance.

\begin{algorithm}
	\caption{Bayesian Optimization} 
        \textbf{Input}  \\
        \hspace*{\algorithmicindent} $\mathcal{D} = \{(X_i, Y_i)|i=1,\ldots, n\}$ : initial dataset \\
        \hspace*{\algorithmicindent} $AF$ : acquisition function
        \begin{algorithmic}[1]
		\For {$iteration=1,2,\ldots$}
        \State Train a surrogate model $Y = \hat{f}(x)$
        \State Select a query $X_{new} \, \leftarrow \mathrm{argmax} \, AF(X)$
		\State Get the observation $Y_{new}$ for $X_{new}$
        \State Update the dataset $\mathcal{D} \leftarrow \mathcal{D} \cup \left ( X_{new}, Y_{new} \right )$
		\EndFor
	\end{algorithmic} 
\end{algorithm}

\subsection*{Surrogate models}
We employed GP regression \cite{rasmussen2003gaussian} to obtain a surrogate model for the MD simulation using a radial basis function kernel as the covariance function. 
The hyperparameters of the covariance function were determined through  maximum likelihood estimation at each step in the BO. 

\subsection*{Calibration}
The values of $C_\text{p}$, refractive index, and Abbe number calculated using MD simulations were calibrated to account for discrepancies in the experimental values using a linear regression model:
\begin{equation}
Y_e = \alpha Y_s + \beta
\end{equation}
where $Y_e$ and $Y_s$ represent the experimental and MD-calculated properties, respectively. The parameters $\alpha$ and $\beta$ were determined using least squares fitting.
Calibration of $C_\text{p}$ was performed using 72 experimental values from PoLyInfo.
The refractive index and Abbe number were calibrated using the experimental properties of 26 polymers extracted from the literature \citep{kawai1995plastic, badur2018high, zhang2023all, suzuki2012synthesis, higashihara2015recent, leosson2012integrated}.

\subsection*{Polymer physical property calculations}

The polymer physical property calculations for the optical polymer design were conducted using RadonPy ver 0.2.5. The chemical structure of a polymer repeating unit, represented by SMILES, was given to RadonPy in addition to the polymerization degree and number of polymer chains forming a simulation cell. Then the following process can be fully automated: (1) conformation search for a monomer with the given repeating unit, (2) atomic charge calculations using the density functional theory (DFT), (3) search for initial configuration of polymer chains (4) assignment of force field parameters using the general Amber force field version 2, (5) generation of isotropic amorphous cells, (6) equilibrium and nonequilibrium MD simulations, and (7) property calculation in the post-processing step. DFT calculations and MD simulations were performed using Psi4 \citep{smith2020psi4} and LAMMPS, respectively, within the RadonPy interface. 

Following the procedure by Hayashi et al. \cite{hayashi2022radonpy}, an amorphous cell containing 10 polymer chains comprising approximately $10,\!000$ atoms was created. The amorphous cell was equilibrated using Larsen's 21-step compression/decompression protocol \citep{larsen2011molecular}, with temperature ascent and descent cycles ranging between 300 and 600 K. Next, NpT simulations were conducted for  5 ns at 300 K and 1 atm, with additional simulations of up to 20 ns if equilibrium was not reached. If equilibrium was still not achieved, the calculations were terminated. 

The refractive index $n$ was derived from the Lorentz--Lorenz equation:
\begin{eqnarray}
\frac{n^2 - 1}{n^2 + 2}= \frac{ 4 \pi}{3}\frac{\rho}{M}\alpha_{\mathrm{polar}}
\end{eqnarray}
Here, $\rho$ is the density from the MD simulation, $\alpha_{\mathrm{polar}}$ is the isotropic dipole polarizability of a repeating unit calculated from the DFT calculation, and $M$ is the molecular weight of a repeating unit. The $\alpha_{\mathrm{polar}}$ was computed by the following procedure: (1) a conformation search of a repeating unit by the protocol implemented in RadonPy, (2) a geometry optimization for the most stable conformer of a repeating unit by the $\omega$B97M–D3BJ functional \citep{mardirossian2016omegab97m, grimme2011effect} combined with the 6–31G(d,p) basis set \citep{ditchfield1971self, francl1982self}, and (3) a single-point polarizability calculation with finite field method using the $\omega$B97M–D3BJ functional combined with the 6–311+G(2d,p) \cite{mardirossian2016omegab97m, grimme2011effect, krishnan1980self, mclean1980contracted, binning1990compact, clark1983efficient, frisch1984self} for H, C, N, O, F, P, S, and Cl atoms, with the 6–311G(d,p) \cite{mardirossian2016omegab97m, grimme2011effect, krishnan1980self, mclean1980contracted, binning1990compact} for Br atom, and with the LanL2DZ basis set \cite{wadt1985ab} for I atom.

The Abbe number $v$ was then calculated using the following equation:
\begin{equation}
v = \frac{n_{589} - 1}{n_{486} - n_{656}},
\end{equation}
where $n_{486}$, $n_{589}$, and $n_{656}$ are the refractive indices at 486, 589, and 656 nm, respectively. The wavelength-dependent refractive indices were also calculated using the Lorentz-Lorenz equation, considering wavelength-dependent polarizability and density. Typically, wavelength-dependent polarizability is calculated using the coupled-perturbed Hartree-Fock method; however, this was not implemented for DFT calculations in Psi4. Therefore, in this study, the wavelength-dependent polarizability $\alpha_{ij}(\omega)$ at frequency $\omega$ was calculated using the sum-over-states approach \cite{rice1991calculation} as follows:
\begin{equation}
\alpha _{ij}(\omega )=2\sum_{n}^{}\left ( \frac{\mu_{i}^{gn}\mu_{j}^{ng}}{\hbar\omega_{gn}-(\hbar\omega )^2/(\hbar\omega_{gn})} \right ),
\end{equation}
where $\mu_{i}^{gn}$ is the transition dipole moment for $i$-axis ($i \in \{ x, y, z\}$) from the ground state $(g)$ to the $n$-th excited state, $\hbar\omega_{gn}$ is the excitation energy from the ground state to the $n$-th excited state, and $\hbar$ is the reduced Planck’s constant.

To calculate the wavelength-dependent polarizability, TD-DFT calculations were performed. In RadonPy, Psi4 is utilized as the quantum chemistry calculation engine. Because TD-DFT calculations are not supported for the $\omega$B97M–D3BJ functional in Psi4, the CAM–B3LYP \cite{yanai2004new}, a GGA functional incorporating important long-range corrections was employed to calculate the excited-states. The 6–311+G(2d,p) basis set was used for H, C, N, O, F, P, S, and Cl atoms, the 6–311G(d,p) was used for Br atom, and the LanL2DZ basis set was used for I atom.

However, the high computational cost makes it impractical to calculate all one-electron excited states using TD-DFT. The tradeoff between computational accuracy and cost was achieved by truncating the number of calculated excited states at a certain point. Preliminary calculations investigated the effect of the number of calculated excited states on the calculated Abbe number accuracy. The left panel of Fig. S1 compares the experimental and calculated Abbe numbers for the 26 polymers obtained by varying $a \in (0.3, 0.01, 0.003, 0.001)$, representing the proportion of excited states considered in the TD-DFT calculation relative to the total excited states. When considering up to 30\% of all excited states ($a = 0.3$), the results agreed closely with the experimental observations, demonstrating the validity of this calculation condition. However, under $a = 0.3$, the computational cost became prohibitive as the molecular size increased, rendering the calculations infeasible. Therefore, further calculations were performed using fewer excited states ($a \in [0.01, 0.001]$). Although the Abbe numbers were underestimated, the correlation coefficient with the experimental values remained at 0.98 for $a = 0.003$ (Fig. S1, right). Hence, we proceeded with the TD-DFT calculations considering 0.3\% of all the excited states ($a = 0.003$).

\subsection*{Experimental validation}

\subsubsection*{Measurements}
$^1$H and $^{13}$C nuclear magnetic resonance (NMR) spectra were recorded using a JEOL JNM-ECS400 (400 MHz) spectrometer with chloroform-$d_{1}$ as the solvent.
Fourier transform infrared (FT-IR) spectra were obtained using a JASCO FT/IR-4100 Fourier transform spectrophotometer.
Size exclusion chromatography (SEC) was performed using a SHIMADZU LC-20AD system equipped with a Shodex RI 501 RI detector and Shodex LF 804 columns. 
The number-average molecular weight ($M_n$) and molecular weight distribution ($M_w/M_n$) were determined via SEC using a polymer/tetrahydrofuran solution at a flow rate of 1.0 mL/min at 40  $^\circ$C calibrated against polystyrene standards.
Thermogravimetric analysis (TGA) was conducted under nitrogen atmosphere using an SII TGA 7300 system. The samples were heated at a rate of 10 $^\circ$C/min within the temperature range of 30--550 $^\circ$C.
The temperature at the 5 \% weight loss (TG$_5$) was determined from the TGA curve.
Differential scanning calorimetry (DSC) measurements were performed under nitrogen flow using an EXSTAR7000 series DSC7020 (Hitachi High Tech) by heating the prepared samples at a rate of 10 $^\circ$C/min.
The glass transition temperature ($T_g$) and melting temperature ($T_m$) were determined from the DSC curves.
The refractive index and extinction coefficient were measured by spectroscopic ellipsometry using an M-2000V-Te  (J. A. Woollam Co.).

\subsubsection*{Reagents}
1,4-butanedithiol, 1,1'-carbonyldiimidazole (CDI), 1,8-diazabicyclo[5.4.0]-7-undecene (DBU), bis(2-mercaptoethyl) sulfide, 1,6-hexanedithiol, 1,4-cyclohexanediol (mixtures of cis and trans isomers), 9,9‐bis(4‐hydroxyphenyl)‐fluorene, 1,4-benzenedimethanethiol and resorcinol were sourced from Tokyo Chemical Industry.
3,6-dioxa-1,8-octanedithiol was sourced from Sigma-Aldrich.
Anhydrous grade solvents, namely, chloroform were purchased from FUJIFILM Wako Pure Chemical Corporation.
All reagents and solvents were used as received.

\subsubsection*{Polymerization}

\noindent
\textbf{Synthesis of P1} \\ 
In a flask, 0.76 g (6.23 mmol) of 1,4-butanedithiol was dissolved in 7 mL of chloroform under a continuous nitrogen flow. Next, 1.06 g (6.54 mmol) of CDI and 1.87 mL of DBU were added sequentially. The solution was stirred at room temperature for 24 h. The crude product was precipitated into a large excess of methanol, filtered, and the residue was dried at 40 $^\circ$C under reduced pressure. \textbf{P1} was obtained as a white solid (0.71 g, 77\% yield).
$M_{n}$ : 20,900, $M_{w} / M_{n}$ : 2.7. $T_g$ : -27 $^\circ$C. $T_m$: 87 $^\circ$C. TG$_5$: 275 $^\circ$C. 
$^1$H NMR (400 MHz, CDCl$_3$, $\delta$, ppm) : 1.66-1.76 (m, 4H, CH$_2$-CH$_2$-CH$_2$), 2.95-3.05 (m, 4H, S-CH$_2$-CH$_2$), 3.82 (s, 6H, O-CH$_3$). 
$^{13}$C NMR (400 MHz, CDCl$_3$, $\delta$, ppm) :  28.7 (CH$_2$-CH$_2$-CH$_2$) , 29.9 (S-CH$_2$-CH$_2$), 189.3 (S-CO-S).
The NMR spectra are shown in Figs. S6 and S7, The SEC curves, IR spectra, TGA curve, and DSC curve are shown in Figs. S8–S11, respectively. The refractive index and extinction coefficient measured using the spectroscopic ellipsometry are shown in Fig. S12.
\\
\\
\noindent
\textbf{Synthesis of P2} \\ 
In a flask, 0.82 g (5.33 mmol) of bis(2-mercaptoethyl) sulfide was dissolved in 7 mL of chloroform under a continuous nitrogen flow. Next, 0.91 g (5.60 mmol) of CDI and 1.60 mL of DBU were added sequentially. The solution was stirred at room temperature for 24 h. The crude product was precipitated into a large excess of methanol and filtered. The residue was dried at 40 $^\circ$C under reduced pressure. \textbf{P2} was obtained as a white solid (0.84 g). This polymer was insoluble in common organic solvents.
\\
\\
\noindent
\textbf{Typical procedure for copolymerization of the raw material of P2 with another monomer} \\ 
In a flask, 1.60 g (10.38 mmol) of bis(2-mercaptoethyl) sulfide and 0.62 g (4.13 mmol) of 1,6-hexanedithiol were dissolved in 19 mL of chloroform under a continuous flow of nitrogen. Next, 2.47 g (15.24 mmol) of CDI and 4.37 mL of DBU were added sequentially. The solution was stirred at room temperature for 24 h. The crude product was precipitated into a large excess of methanol and filtered. The residue was dried at 40 $^\circ$C under reduced pressure to obtain a white solid  (2.07 g) .

\section*{Data availability}
The experimental and computational datasets are available at GitHub \url{https://github.com/s-nanjo/Spacier/tree/main/Optical_Polymer_Dataset}. 

\section*{Code availability}
The source code of SPACIER is available from Github \url{https://github.com/s-nanjo/Spacier/}. 

\section*{Author contributions}

R.Y. and S.N. conceptualized and outlined the project and provided its main ideas. S.N., along with R.Y. and Y.H., implemented the core machine-learning algorithms and conducted experiments. S.N. and A. developed the SPACIER software. Y. H. created the RadonPy workflow for the Abbe number. S.N. with assistance from H.M., K.H-S., R.H., and T.H. synthesized the polymers and assessed their properties. S.N. and R.Y. wrote the manuscript.

\section*{Acknowledgments}\label{sec5}
We express our sincere gratitude to Professor Shinji Ando at Tokyo Institute of Technology for his valuable contributions to the discussions of this study.
This research received support from MEXT as ``Program for Promoting Researches on the Supercomputer Fugaku'' (project ID: hp210264), JST CREST (Grant Numbers JPMJCR19I3, JPMJCR22O3, JPMJCR2332), MEXT/JSPS KAKENHI Grant-in-Aid for Scientific Research on Innovative Areas (19H05820), Grant-in-Aid for Scientific Research (A) (19H01132), and Grant-in-Aid for Scientific Research (C) (22K11949). Computational resources were provided by Fugaku at the RIKEN Center for Computational Science, Kobe, Japan (hp210264) and the supercomputer at the Research Center for Computational Science, Okazaki, Japan (project: 23-IMS-C113, 24-IMS-C107).

\clearpage
\bibliography{sn-bibliography}


\begin{thebibliography}{59}
\ifx \bisbn   \undefined \def \bisbn  #1{ISBN #1}\fi
\ifx \binits  \undefined \def \binits#1{#1}\fi
\ifx \bauthor  \undefined \def \bauthor#1{#1}\fi
\ifx \batitle  \undefined \def \batitle#1{#1}\fi
\ifx \bjtitle  \undefined \def \bjtitle#1{#1}\fi
\ifx \bvolume  \undefined \def \bvolume#1{\textbf{#1}}\fi
\ifx \byear  \undefined \def \byear#1{#1}\fi
\ifx \bissue  \undefined \def \bissue#1{#1}\fi
\ifx \bfpage  \undefined \def \bfpage#1{#1}\fi
\ifx \blpage  \undefined \def \blpage #1{#1}\fi
\ifx \burl  \undefined \def \burl#1{\textsf{#1}}\fi
\ifx \doiurl  \undefined \def \doiurl#1{\url{https://doi.org/#1}}\fi
\ifx \betal  \undefined \def \betal{\textit{et al.}}\fi
\ifx \binstitute  \undefined \def \binstitute#1{#1}\fi
\ifx \binstitutionaled  \undefined \def \binstitutionaled#1{#1}\fi
\ifx \bctitle  \undefined \def \bctitle#1{#1}\fi
\ifx \beditor  \undefined \def \beditor#1{#1}\fi
\ifx \bpublisher  \undefined \def \bpublisher#1{#1}\fi
\ifx \bbtitle  \undefined \def \bbtitle#1{#1}\fi
\ifx \bedition  \undefined \def \bedition#1{#1}\fi
\ifx \bseriesno  \undefined \def \bseriesno#1{#1}\fi
\ifx \blocation  \undefined \def \blocation#1{#1}\fi
\ifx \bsertitle  \undefined \def \bsertitle#1{#1}\fi
\ifx \bsnm \undefined \def \bsnm#1{#1}\fi
\ifx \bsuffix \undefined \def \bsuffix#1{#1}\fi
\ifx \bparticle \undefined \def \bparticle#1{#1}\fi
\ifx \barticle \undefined \def \barticle#1{#1}\fi
\bibcommenthead
\ifx \bconfdate \undefined \def \bconfdate #1{#1}\fi
\ifx \botherref \undefined \def \botherref #1{#1}\fi
\ifx \url \undefined \def \url#1{\textsf{#1}}\fi
\ifx \bchapter \undefined \def \bchapter#1{#1}\fi
\ifx \bbook \undefined \def \bbook#1{#1}\fi
\ifx \bcomment \undefined \def \bcomment#1{#1}\fi
\ifx \oauthor \undefined \def \oauthor#1{#1}\fi
\ifx \citeauthoryear \undefined \def \citeauthoryear#1{#1}\fi
\ifx \endbibitem  \undefined \def \endbibitem {}\fi
\ifx \bconflocation  \undefined \def \bconflocation#1{#1}\fi
\ifx \arxivurl  \undefined \def \arxivurl#1{\textsf{#1}}\fi
\csname PreBibitemsHook\endcsname

%%% 1
\bibitem[\protect\citeauthoryear{Agrawal and Choudhary}{2016}]{agrawal2016perspective}
\begin{botherref}
\oauthor{\bsnm{Agrawal}, \binits{A.}},
\oauthor{\bsnm{Choudhary}, \binits{A.}}:
Perspective: materials informatics and big data: realization of the “fourth paradigm” of science in materials science.
APL Materials
\textbf{4}(5)
(2016)
\end{botherref}
\endbibitem

%%% 2
\bibitem[\protect\citeauthoryear{Wu et~al.}{2019}]{wu2019machine}
\begin{barticle}
\bauthor{\bsnm{Wu}, \binits{S.}},
\bauthor{\bsnm{Kondo}, \binits{Y.}},
\bauthor{\bsnm{Kakimoto}, \binits{M.-A.}},
\bauthor{\bsnm{Yang}, \binits{B.}},
\bauthor{\bsnm{Yamada}, \binits{H.}},
\bauthor{\bsnm{Kuwajima}, \binits{I.}},
\bauthor{\bsnm{Lambard}, \binits{G.}},
\bauthor{\bsnm{Hongo}, \binits{K.}},
\bauthor{\bsnm{Xu}, \binits{Y.}},
\bauthor{\bsnm{Shiomi}, \binits{J.}},
\bauthor{\bsnm{Schick}, \binits{C.}},
\bauthor{\bsnm{Morikawa}, \binits{J.}},
\bauthor{\bsnm{Yoshida}, \binits{R.}}:
\batitle{Machine-learning-assisted discovery of polymers with high thermal conductivity using a molecular design algorithm}.
\bjtitle{npj {C}omputational {M}aterials}
\bvolume{5}(\bissue{1}),
\bfpage{66}
(\byear{2019})
\end{barticle}
\endbibitem

%%% 3
\bibitem[\protect\citeauthoryear{Merchant et~al.}{2023}]{merchant2023scaling}
\begin{barticle}
\bauthor{\bsnm{Merchant}, \binits{A.}},
\bauthor{\bsnm{Batzner}, \binits{S.}},
\bauthor{\bsnm{Schoenholz}, \binits{S.S.}},
\bauthor{\bsnm{Aykol}, \binits{M.}},
\bauthor{\bsnm{Cheon}, \binits{G.}},
\bauthor{\bsnm{Cubuk}, \binits{E.D.}}:
\batitle{Scaling deep learning for materials discovery}.
\bjtitle{Nature}
\bvolume{624}(\bissue{7990}),
\bfpage{80}--\blpage{85}
(\byear{2023})
\end{barticle}
\endbibitem

%%% 4
\bibitem[\protect\citeauthoryear{Szymanski et~al.}{2023}]{szymanski2023autonomous}
\begin{barticle}
\bauthor{\bsnm{Szymanski}, \binits{N.J.}},
\bauthor{\bsnm{Rendy}, \binits{B.}},
\bauthor{\bsnm{Fei}, \binits{Y.}},
\bauthor{\bsnm{Kumar}, \binits{R.E.}},
\bauthor{\bsnm{He}, \binits{T.}},
\bauthor{\bsnm{Milsted}, \binits{D.}},
\bauthor{\bsnm{McDermott}, \binits{M.J.}},
\bauthor{\bsnm{Gallant}, \binits{M.}},
\bauthor{\bsnm{Cubuk}, \binits{E.D.}},
\bauthor{\bsnm{Merchant}, \binits{A.}},
\bauthor{\bsnm{Kim}, \binits{H.}},
\bauthor{\bsnm{Jain}, \binits{A.}},
\bauthor{\bsnm{Bartel}, \binits{C.J.}},
\bauthor{\bsnm{Persson}, \binits{K.}},
\bauthor{\bsnm{Zeng}, \binits{Y.}},
\bauthor{\bsnm{Ceder}, \binits{G.}}:
\batitle{An autonomous laboratory for the accelerated synthesis of novel materials}.
\bjtitle{Nature}
\bvolume{624}(\bissue{7990}),
\bfpage{86}--\blpage{91}
(\byear{2023})
\end{barticle}
\endbibitem

%%% 5
\bibitem[\protect\citeauthoryear{Rao et~al.}{2022}]{rao2022machine}
\begin{barticle}
\bauthor{\bsnm{Rao}, \binits{Z.}},
\bauthor{\bsnm{Tung}, \binits{P.-Y.}},
\bauthor{\bsnm{Xie}, \binits{R.}},
\bauthor{\bsnm{Wei}, \binits{Y.}},
\bauthor{\bsnm{Zhang}, \binits{H.}},
\bauthor{\bsnm{Ferrari}, \binits{A.}},
\bauthor{\bsnm{Klaver}, \binits{T.P.C.}},
\bauthor{\bsnm{K\"{o}rmann}, \binits{F.}},
\bauthor{\bsnm{Sukumar}, \binits{P.T.}},
\bauthor{\bsnm{Silva}, \binits{A.}},
\bauthor{\bsnm{Chen}, \binits{Y.}},
\bauthor{\bsnm{Li}, \binits{Z.}},
\bauthor{\bsnm{Ponge}, \binits{D.}},
\bauthor{\bsnm{Neugebauer}, \binits{J.}},
\bauthor{\bsnm{Gutfleisch}, \binits{O.}},
\bauthor{\bsnm{Bauer}, \binits{S.}},
\bauthor{\bsnm{Raabe}, \binits{D.}}:
\batitle{Machine learning--enabled high-entropy alloy discovery}.
\bjtitle{Science}
\bvolume{378}(\bissue{6615}),
\bfpage{78}--\blpage{85}
(\byear{2022})
\end{barticle}
\endbibitem

%%% 6
\bibitem[\protect\citeauthoryear{Zhong et~al.}{2020}]{zhong2020accelerated}
\begin{barticle}
\bauthor{\bsnm{Zhong}, \binits{M.}},
\bauthor{\bsnm{Tran}, \binits{K.}},
\bauthor{\bsnm{Min}, \binits{Y.}},
\bauthor{\bsnm{Wang}, \binits{C.}},
\bauthor{\bsnm{Wang}, \binits{Z.}},
\bauthor{\bsnm{Dinh}, \binits{C.-T.}},
\bauthor{\bsnm{De~Luna}, \binits{P.}},
\bauthor{\bsnm{Yu}, \binits{Z.}},
\bauthor{\bsnm{Rasouli}, \binits{A.S.}},
\bauthor{\bsnm{Brodersen}, \binits{P.}},
\bauthor{\bsnm{Sun}, \binits{S.}},
\bauthor{\bsnm{Voznyy}, \binits{O.}},
\bauthor{\bsnm{Tan}, \binits{C.-S.}},
\bauthor{\bsnm{Askerka}, \binits{M.}},
\bauthor{\bsnm{Che}, \binits{F.}},
\bauthor{\bsnm{Liu}, \binits{M.}},
\bauthor{\bsnm{Seifitokaldani}, \binits{A.}},
\bauthor{\bsnm{Pang}, \binits{Y.}},
\bauthor{\bsnm{Lo}, \binits{S.-C.}},
\bauthor{\bsnm{Ip}, \binits{A.}},
\bauthor{\bsnm{Ulissi}, \binits{Z.}},
\bauthor{\bsnm{Sargent}, \binits{E.H.}}:
\batitle{Accelerated discovery of {CO}2 electrocatalysts using active machine learning}.
\bjtitle{Nature}
\bvolume{581}(\bissue{7807}),
\bfpage{178}--\blpage{183}
(\byear{2020})
\end{barticle}
\endbibitem

%%% 7
\bibitem[\protect\citeauthoryear{Kim et~al.}{2020}]{kim2019artificial}
\begin{barticle}
\bauthor{\bsnm{Kim}, \binits{M.}},
\bauthor{\bsnm{Yeo}, \binits{B.C.}},
\bauthor{\bsnm{Park}, \binits{Y.}},
\bauthor{\bsnm{Lee}, \binits{H.M.}},
\bauthor{\bsnm{Han}, \binits{S.S.}},
\bauthor{\bsnm{Kim}, \binits{D.}}:
\batitle{Artificial intelligence to accelerate the discovery of {N}2 electroreduction catalysts}.
\bjtitle{Chemistry of Materials}
\bvolume{32}(\bissue{2}),
\bfpage{709}--\blpage{720}
(\byear{2020})
\end{barticle}
\endbibitem

%%% 8
\bibitem[\protect\citeauthoryear{Liu et~al.}{2021}]{liu2021machine}
\begin{barticle}
\bauthor{\bsnm{Liu}, \binits{C.}},
\bauthor{\bsnm{Fujita}, \binits{E.}},
\bauthor{\bsnm{Katsura}, \binits{Y.}},
\bauthor{\bsnm{Inada}, \binits{Y.}},
\bauthor{\bsnm{Ishikawa}, \binits{A.}},
\bauthor{\bsnm{Tamura}, \binits{R.}},
\bauthor{\bsnm{Kimura}, \binits{K.}},
\bauthor{\bsnm{Yoshida}, \binits{R.}}:
\batitle{Machine learning to predict quasicrystals from chemical compositions}.
\bjtitle{Advanced Materials}
\bvolume{33}(\bissue{36}),
\bfpage{2102507}
(\byear{2021})
\end{barticle}
\endbibitem

%%% 9
\bibitem[\protect\citeauthoryear{Liu et~al.}{2023}]{liu2023quasicrystals}
\begin{barticle}
\bauthor{\bsnm{Liu}, \binits{C.}},
\bauthor{\bsnm{Kitahara}, \binits{K.}},
\bauthor{\bsnm{Ishikawa}, \binits{A.}},
\bauthor{\bsnm{Hiroto}, \binits{T.}},
\bauthor{\bsnm{Singh}, \binits{A.}},
\bauthor{\bsnm{Fujita}, \binits{E.}},
\bauthor{\bsnm{Katsura}, \binits{Y.}},
\bauthor{\bsnm{Inada}, \binits{Y.}},
\bauthor{\bsnm{Tamura}, \binits{R.}},
\bauthor{\bsnm{Kimura}, \binits{K.}},
\bauthor{\bsnm{Yoshida}, \binits{R.}}:
\batitle{Quasicrystals predicted and discovered by machine learning}.
\bjtitle{Physical Review Materials}
\bvolume{7}(\bissue{9}),
\bfpage{093805}
(\byear{2023})
\end{barticle}
\endbibitem

%%% 10
\bibitem[\protect\citeauthoryear{Uryu et~al.}{2024}]{uryu2024deep}
\begin{barticle}
\bauthor{\bsnm{Uryu}, \binits{H.}},
\bauthor{\bsnm{Yamada}, \binits{T.}},
\bauthor{\bsnm{Kitahara}, \binits{K.}},
\bauthor{\bsnm{Singh}, \binits{A.}},
\bauthor{\bsnm{Iwasaki}, \binits{Y.}},
\bauthor{\bsnm{Kimura}, \binits{K.}},
\bauthor{\bsnm{Hiroki}, \binits{K.}},
\bauthor{\bsnm{Miyao}, \binits{N.}},
\bauthor{\bsnm{Ishikawa}, \binits{A.}},
\bauthor{\bsnm{Tamura}, \binits{R.}},
\bauthor{\bsnm{Ohhashi}, \binits{S.}},
\bauthor{\bsnm{Liu}, \binits{C.}},
\bauthor{\bsnm{Yoshida}, \binits{R.}}:
\batitle{Deep learning enables rapid identification of a new quasicrystal from multiphase powder diffraction patterns}.
\bjtitle{Advanced Science}
\bvolume{11}(\bissue{1}),
\bfpage{2304546}
(\byear{2024})
\end{barticle}
\endbibitem

%%% 11
\bibitem[\protect\citeauthoryear{Ishii et~al.}{2024}]{ishii2024nims}
\begin{barticle}
\bauthor{\bsnm{Ishii}, \binits{M.}},
\bauthor{\bsnm{Ito}, \binits{T.}},
\bauthor{\bsnm{Sado}, \binits{H.}},
\bauthor{\bsnm{Kuwajima}, \binits{I.}}:
\batitle{{NIMS} polymer database {PoLyInfo} ({I}): an overarching view of half a million data points}.
\bjtitle{Science and Technology of Advanced Materials: Methods}
\bvolume{4}(\bissue{1}),
\bfpage{2354649}
(\byear{2024})
\end{barticle}
\endbibitem

%%% 12
\bibitem[\protect\citeauthoryear{Kim et~al.}{2018}]{kim2018polymer}
\begin{barticle}
\bauthor{\bsnm{Kim}, \binits{C.}},
\bauthor{\bsnm{Chandrasekaran}, \binits{A.}},
\bauthor{\bsnm{Huan}, \binits{T.D.}},
\bauthor{\bsnm{Das}, \binits{D.}},
\bauthor{\bsnm{Ramprasad}, \binits{R.}}:
\batitle{Polymer genome: a data-powered polymer informatics platform for property predictions}.
\bjtitle{The Journal of Physical Chemistry C}
\bvolume{122}(\bissue{31}),
\bfpage{17575}--\blpage{17585}
(\byear{2018})
\end{barticle}
\endbibitem

%%% 13
\bibitem[\protect\citeauthoryear{Takahashi et~al.}{2024}]{takahashi2024copoldb}
\begin{barticle}
\bauthor{\bsnm{Takahashi}, \binits{K.-i.}},
\bauthor{\bsnm{Mamitsuka}, \binits{H.}},
\bauthor{\bsnm{Tosaka}, \binits{M.}},
\bauthor{\bsnm{Zhu}, \binits{N.}},
\bauthor{\bsnm{Yamago}, \binits{S.}}:
\batitle{Co{P}ol{DB}: a copolymerization database for radical polymerization}.
\bjtitle{Polymer Chemistry}
\bvolume{15}(\bissue{10}),
\bfpage{965}--\blpage{971}
(\byear{2024})
\end{barticle}
\endbibitem

%%% 14
\bibitem[\protect\citeauthoryear{Brochu et~al.}{2010}]{brochu2010tutorial}
\begin{botherref}
\oauthor{\bsnm{Brochu}, \binits{E.}},
\oauthor{\bsnm{Cora}, \binits{V.M.}},
\oauthor{\bsnm{De~Freitas}, \binits{N.}}:
A tutorial on bayesian optimization of expensive cost functions, with application to active user modeling and hierarchical reinforcement learning.
arXiv preprint arXiv:1012.2599
(2010)
\end{botherref}
\endbibitem

%%% 15
\bibitem[\protect\citeauthoryear{Shahriari et~al.}{2015}]{shahriari2015taking}
\begin{barticle}
\bauthor{\bsnm{Shahriari}, \binits{B.}},
\bauthor{\bsnm{Swersky}, \binits{K.}},
\bauthor{\bsnm{Wang}, \binits{Z.}},
\bauthor{\bsnm{Adams}, \binits{R.P.}},
\bauthor{\bsnm{De~Freitas}, \binits{N.}}:
\batitle{Taking the human out of the loop: a review of bayesian optimization}.
\bjtitle{Proceedings of the IEEE}
\bvolume{104}(\bissue{1}),
\bfpage{148}--\blpage{175}
(\byear{2015})
\end{barticle}
\endbibitem

%%% 16
\bibitem[\protect\citeauthoryear{Frazier}{2018}]{frazier2018tutorial}
\begin{botherref}
\oauthor{\bsnm{Frazier}, \binits{P.I.}}:
A tutorial on bayesian optimization.
arXiv preprint arXiv:1807.02811
(2018)
\end{botherref}
\endbibitem

%%% 17
\bibitem[\protect\citeauthoryear{Seko et~al.}{2015}]{seko2015prediction}
\begin{barticle}
\bauthor{\bsnm{Seko}, \binits{A.}},
\bauthor{\bsnm{Togo}, \binits{A.}},
\bauthor{\bsnm{Hayashi}, \binits{H.}},
\bauthor{\bsnm{Tsuda}, \binits{K.}},
\bauthor{\bsnm{Chaput}, \binits{L.}},
\bauthor{\bsnm{Tanaka}, \binits{I.}}:
\batitle{Prediction of low-thermal-conductivity compounds with first-principles anharmonic lattice-dynamics calculations and bayesian optimization}.
\bjtitle{Physical Review Letters}
\bvolume{115}(\bissue{20}),
\bfpage{205901}
(\byear{2015})
\end{barticle}
\endbibitem

%%% 18
\bibitem[\protect\citeauthoryear{Ju et~al.}{2017}]{ju2017designing}
\begin{barticle}
\bauthor{\bsnm{Ju}, \binits{S.}},
\bauthor{\bsnm{Shiga}, \binits{T.}},
\bauthor{\bsnm{Feng}, \binits{L.}},
\bauthor{\bsnm{Hou}, \binits{Z.}},
\bauthor{\bsnm{Tsuda}, \binits{K.}},
\bauthor{\bsnm{Shiomi}, \binits{J.}}:
\batitle{Designing nanostructures for phonon transport via bayesian optimization}.
\bjtitle{Physical Review X}
\bvolume{7}(\bissue{2}),
\bfpage{021024}
(\byear{2017})
\end{barticle}
\endbibitem

%%% 19
\bibitem[\protect\citeauthoryear{Yamashita et~al.}{2021}]{yamashita2021cryspy}
\begin{barticle}
\bauthor{\bsnm{Yamashita}, \binits{T.}},
\bauthor{\bsnm{Kanehira}, \binits{S.}},
\bauthor{\bsnm{Sato}, \binits{N.}},
\bauthor{\bsnm{Kino}, \binits{H.}},
\bauthor{\bsnm{Terayama}, \binits{K.}},
\bauthor{\bsnm{Sawahata}, \binits{H.}},
\bauthor{\bsnm{Sato}, \binits{T.}},
\bauthor{\bsnm{Utsuno}, \binits{F.}},
\bauthor{\bsnm{Tsuda}, \binits{K.}},
\bauthor{\bsnm{Miyake}, \binits{T.}},
\bauthor{\bsnm{Oguchi}, \binits{T.}}:
\batitle{Cry{SPY}: a crystal structure prediction tool accelerated by machine learning}.
\bjtitle{Science and Technology of Advanced Materials: Methods}
\bvolume{1}(\bissue{1}),
\bfpage{87}--\blpage{97}
(\byear{2021})
\end{barticle}
\endbibitem

%%% 20
\bibitem[\protect\citeauthoryear{Tran et~al.}{2019}]{tran2019pbo}
\begin{barticle}
\bauthor{\bsnm{Tran}, \binits{A.}},
\bauthor{\bsnm{Sun}, \binits{J.}},
\bauthor{\bsnm{Furlan}, \binits{J.M.}},
\bauthor{\bsnm{Pagalthivarthi}, \binits{K.V.}},
\bauthor{\bsnm{Visintainer}, \binits{R.J.}},
\bauthor{\bsnm{Wang}, \binits{Y.}}:
\batitle{{pBO-2GP-3B}: a batch parallel known/unknown constrained bayesian optimization with feasibility classification and its applications in computational fluid dynamics}.
\bjtitle{Computer Methods in Applied Mechanics and Engineering}
\bvolume{347},
\bfpage{827}--\blpage{852}
(\byear{2019})
\end{barticle}
\endbibitem

%%% 21
\bibitem[\protect\citeauthoryear{Sakurai et~al.}{2019}]{sakurai2019ultranarrow}
\begin{barticle}
\bauthor{\bsnm{Sakurai}, \binits{A.}},
\bauthor{\bsnm{Yada}, \binits{K.}},
\bauthor{\bsnm{Simomura}, \binits{T.}},
\bauthor{\bsnm{Ju}, \binits{S.}},
\bauthor{\bsnm{Kashiwagi}, \binits{M.}},
\bauthor{\bsnm{Okada}, \binits{H.}},
\bauthor{\bsnm{Nagao}, \binits{T.}},
\bauthor{\bsnm{Tsuda}, \binits{K.}},
\bauthor{\bsnm{Shiomi}, \binits{J.}}:
\batitle{Ultranarrow-band wavelength-selective thermal emission with aperiodic multilayered metamaterials designed by bayesian optimization}.
\bjtitle{ACS Central Science}
\bvolume{5}(\bissue{2}),
\bfpage{319}--\blpage{326}
(\byear{2019})
\end{barticle}
\endbibitem

%%% 22
\bibitem[\protect\citeauthoryear{Sumita et~al.}{2018}]{sumita2018hunting}
\begin{barticle}
\bauthor{\bsnm{Sumita}, \binits{M.}},
\bauthor{\bsnm{Yang}, \binits{X.}},
\bauthor{\bsnm{Ishihara}, \binits{S.}},
\bauthor{\bsnm{Tamura}, \binits{R.}},
\bauthor{\bsnm{Tsuda}, \binits{K.}}:
\batitle{Hunting for organic molecules with artificial intelligence: molecules optimized for desired excitation energies}.
\bjtitle{ACS Central Science}
\bvolume{4}(\bissue{9}),
\bfpage{1126}--\blpage{1133}
(\byear{2018})
\end{barticle}
\endbibitem

%%% 23
\bibitem[\protect\citeauthoryear{Wang et~al.}{2020}]{wang2020toward}
\begin{barticle}
\bauthor{\bsnm{Wang}, \binits{Y.}},
\bauthor{\bsnm{Xie}, \binits{T.}},
\bauthor{\bsnm{France-Lanord}, \binits{A.}},
\bauthor{\bsnm{Berkley}, \binits{A.}},
\bauthor{\bsnm{Johnson}, \binits{J.A.}},
\bauthor{\bsnm{Shao-Horn}, \binits{Y.}},
\bauthor{\bsnm{Grossman}, \binits{J.C.}}:
\batitle{Toward designing highly conductive polymer electrolytes by machine learning assisted coarse-grained molecular dynamics}.
\bjtitle{Chemistry of Materials}
\bvolume{32}(\bissue{10}),
\bfpage{4144}--\blpage{4151}
(\byear{2020})
\end{barticle}
\endbibitem

%%% 24
\bibitem[\protect\citeauthoryear{Wu and Zhang}{2023}]{wu2023coarse}
\begin{barticle}
\bauthor{\bsnm{Wu}, \binits{T.}},
\bauthor{\bsnm{Zhang}, \binits{P.}}:
\batitle{Coarse-grained simulation of {PEO/LiTFSI} electrolytes with assistance of bayesian optimization}.
\bjtitle{Macromolecules}
\bvolume{56}(\bissue{17}),
\bfpage{6609}--\blpage{6617}
(\byear{2023})
\end{barticle}
\endbibitem

%%% 25
\bibitem[\protect\citeauthoryear{Hayashi et~al.}{2022}]{hayashi2022radonpy}
\begin{barticle}
\bauthor{\bsnm{Hayashi}, \binits{Y.}},
\bauthor{\bsnm{Shiomi}, \binits{J.}},
\bauthor{\bsnm{Morikawa}, \binits{J.}},
\bauthor{\bsnm{Yoshida}, \binits{R.}}:
\batitle{Radon{P}y: automated physical property calculation using all-atom classical molecular dynamics simulations for polymer informatics}.
\bjtitle{npj Computational Materials}
\bvolume{8}(\bissue{1}),
\bfpage{222}
(\byear{2022})
\end{barticle}
\endbibitem

%%% 26
\bibitem[\protect\citeauthoryear{Ohno et~al.}{2023}]{ohno2023smipoly}
\begin{barticle}
\bauthor{\bsnm{Ohno}, \binits{M.}},
\bauthor{\bsnm{Hayashi}, \binits{Y.}},
\bauthor{\bsnm{Zhang}, \binits{Q.}},
\bauthor{\bsnm{Kaneko}, \binits{Y.}},
\bauthor{\bsnm{Yoshida}, \binits{R.}}:
\batitle{{SMiPoly}: generation of a synthesizable polymer virtual library using rule-based polymerization reactions}.
\bjtitle{Journal of Chemical Information and Modeling}
\bvolume{63}(\bissue{17}),
\bfpage{5539}--\blpage{5548}
(\byear{2023})
\end{barticle}
\endbibitem

%%% 27
\bibitem[\protect\citeauthoryear{Kusaba et~al.}{2023}]{kusaba2023representation}
\begin{barticle}
\bauthor{\bsnm{Kusaba}, \binits{M.}},
\bauthor{\bsnm{Hayashi}, \binits{Y.}},
\bauthor{\bsnm{Liu}, \binits{C.}},
\bauthor{\bsnm{Wakiuchi}, \binits{A.}},
\bauthor{\bsnm{Yoshida}, \binits{R.}}:
\batitle{Representation of materials by kernel mean embedding}.
\bjtitle{Physical Review B}
\bvolume{108}(\bissue{13}),
\bfpage{134107}
(\byear{2023})
\end{barticle}
\endbibitem

%%% 28
\bibitem[\protect\citeauthoryear{Yang et~al.}{2019}]{yang2019multi}
\begin{barticle}
\bauthor{\bsnm{Yang}, \binits{K.}},
\bauthor{\bsnm{Emmerich}, \binits{M.}},
\bauthor{\bsnm{Deutz}, \binits{A.}},
\bauthor{\bsnm{B{\"a}ck}, \binits{T.}}:
\batitle{Multi-objective bayesian global optimization using expected hypervolume improvement gradient}.
\bjtitle{Swarm and Evolutionary Computation}
\bvolume{44},
\bfpage{945}--\blpage{956}
(\byear{2019})
\end{barticle}
\endbibitem

%%% 29
\bibitem[\protect\citeauthoryear{Okutsu et~al.}{2008}]{okutsu2008sulfur}
\begin{barticle}
\bauthor{\bsnm{Okutsu}, \binits{R.}},
\bauthor{\bsnm{Ando}, \binits{S.}},
\bauthor{\bsnm{Ueda}, \binits{M.}}:
\batitle{Sulfur-containing poly (meth) acrylates with high refractive indices and high abbe’s numbers}.
\bjtitle{Chemistry of Materials}
\bvolume{20}(\bissue{12}),
\bfpage{4017}--\blpage{4023}
(\byear{2008})
\end{barticle}
\endbibitem

%%% 30
\bibitem[\protect\citeauthoryear{Cai et~al.}{2015}]{cai2015sulfonyl}
\begin{barticle}
\bauthor{\bsnm{Cai}, \binits{B.}},
\bauthor{\bsnm{Kaino}, \binits{T.}},
\bauthor{\bsnm{Sugihara}, \binits{O.}}:
\batitle{Sulfonyl-containing polymer and its alumina nanocomposite with high abbe number and high refractive index}.
\bjtitle{Optical Materials Express}
\bvolume{5}(\bissue{5}),
\bfpage{1210}--\blpage{1216}
(\byear{2015})
\end{barticle}
\endbibitem

%%% 31
\bibitem[\protect\citeauthoryear{Suzuki et~al.}{2012}]{suzuki2012synthesis}
\begin{barticle}
\bauthor{\bsnm{Suzuki}, \binits{Y.}},
\bauthor{\bsnm{Higashihara}, \binits{T.}},
\bauthor{\bsnm{Ando}, \binits{S.}},
\bauthor{\bsnm{Ueda}, \binits{M.}}:
\batitle{Synthesis and characterization of high refractive index and high abbe’s number poly (thioether sulfone) s based on tricyclo [5.2. 1.02, 6] decane moiety}.
\bjtitle{Macromolecules}
\bvolume{45}(\bissue{8}),
\bfpage{3402}--\blpage{3408}
(\byear{2012})
\end{barticle}
\endbibitem

%%% 32
\bibitem[\protect\citeauthoryear{Berti et~al.}{1988}]{berti1988sulfur}
\begin{barticle}
\bauthor{\bsnm{Berti}, \binits{C.}},
\bauthor{\bsnm{Marianucci}, \binits{E.}},
\bauthor{\bsnm{Pilati}, \binits{F.}}:
\batitle{Sulfur-containing polymers, 4. polymers with thiocarbonate and dithiocarbonate moieties from aliphatic dithiols. syntheses and characterization}.
\bjtitle{Die Makromolekulare Chemie}
\bvolume{189}(\bissue{6}),
\bfpage{1323}--\blpage{1330}
(\byear{1988})
\end{barticle}
\endbibitem

%%% 33
\bibitem[\protect\citeauthoryear{Wnuczek et~al.}{2021}]{wnuczek2021synthesis}
\begin{barticle}
\bauthor{\bsnm{Wnuczek}, \binits{K.}},
\bauthor{\bsnm{Puszka}, \binits{A.}},
\bauthor{\bsnm{Podko{\'s}cielna}, \binits{B.}}:
\batitle{Synthesis and spectroscopic analyses of new polycarbonates based on bisphenol {A}-free components}.
\bjtitle{Polymers}
\bvolume{13}(\bissue{24}),
\bfpage{4437}
(\byear{2021})
\end{barticle}
\endbibitem

%%% 34
\bibitem[\protect\citeauthoryear{Sehn et~al.}{2022}]{sehn2022straightforward}
\begin{barticle}
\bauthor{\bsnm{Sehn}, \binits{T.}},
\bauthor{\bsnm{Huber}, \binits{B.}},
\bauthor{\bsnm{Fanelli}, \binits{J.}},
\bauthor{\bsnm{Mutlu}, \binits{H.}}:
\batitle{Straightforward synthesis of aliphatic polydithiocarbonates from commercially available starting materials}.
\bjtitle{Polymer Chemistry}
\bvolume{13}(\bissue{42}),
\bfpage{5965}--\blpage{5973}
(\byear{2022})
\end{barticle}
\endbibitem

%%% 35
\bibitem[\protect\citeauthoryear{Yoshida and Endo}{2018}]{yoshida2018synthesis}
\begin{barticle}
\bauthor{\bsnm{Yoshida}, \binits{Y.}},
\bauthor{\bsnm{Endo}, \binits{T.}}:
\batitle{Synthesis of polydithiourethanes and their thermal, optical, and mechanical properties originated from monomers structure}.
\bjtitle{Journal of Polymer Science Part A: Polymer Chemistry}
\bvolume{56}(\bissue{19}),
\bfpage{2255}--\blpage{2262}
(\byear{2018})
\end{barticle}
\endbibitem

%%% 36
\bibitem[\protect\citeauthoryear{Watanabe et~al.}{2024}]{watanabe2024polarizable}
\begin{botherref}
\oauthor{\bsnm{Watanabe}, \binits{S.}},
\oauthor{\bsnm{Cavinato}, \binits{L.M.}},
\oauthor{\bsnm{Calvi}, \binits{V.}},
\oauthor{\bsnm{Rijn}, \binits{R.}},
\oauthor{\bsnm{Costa}, \binits{R.D.}},
\oauthor{\bsnm{Oyaizu}, \binits{K.}}:
Polarizable {H-Bond} concept in aromatic poly (thiourea) s: unprecedented high refractive index, transmittance, and degradability at force to enhance lighting efficiency.
Advanced Functional Materials,
2404433
(2024)
\end{botherref}
\endbibitem

%%% 37
\bibitem[\protect\citeauthoryear{Oyama et~al.}{1999}]{oyama1999polymer}
\begin{barticle}
\bauthor{\bsnm{Oyama}, \binits{T.}},
\bauthor{\bsnm{Naka}, \binits{K.}},
\bauthor{\bsnm{Chujo}, \binits{Y.}}:
\batitle{Polymer homologue of {DMSO}: synthesis of poly (ethylene sulfoxide) by selective oxidation of poly (ethylene sulfide)}.
\bjtitle{Macromolecules}
\bvolume{32}(\bissue{16}),
\bfpage{5240}--\blpage{5242}
(\byear{1999})
\end{barticle}
\endbibitem

%%% 38
\bibitem[\protect\citeauthoryear{Aoki et~al.}{2023}]{aoki2023multitask}
\begin{barticle}
\bauthor{\bsnm{Aoki}, \binits{Y.}},
\bauthor{\bsnm{Wu}, \binits{S.}},
\bauthor{\bsnm{Tsurimoto}, \binits{T.}},
\bauthor{\bsnm{Hayashi}, \binits{Y.}},
\bauthor{\bsnm{Minami}, \binits{S.}},
\bauthor{\bsnm{Tadamichi}, \binits{O.}},
\bauthor{\bsnm{Shiratori}, \binits{K.}},
\bauthor{\bsnm{Yoshida}, \binits{R.}}:
\batitle{Multitask machine learning to predict polymer--solvent miscibility using {Flory--Huggins} interaction parameters}.
\bjtitle{Macromolecules}
\bvolume{56}(\bissue{14}),
\bfpage{5446}--\blpage{5456}
(\byear{2023})
\end{barticle}
\endbibitem

%%% 39
\bibitem[\protect\citeauthoryear{Weininger}{1988}]{weininger1988smiles}
\begin{barticle}
\bauthor{\bsnm{Weininger}, \binits{D.}}:
\batitle{{SMILES}, a chemical language and information system. 1. {I}ntroduction to methodology and encoding rules}.
\bjtitle{Journal of Chemical Information and Computer Sciences}
\bvolume{28}(\bissue{1}),
\bfpage{31}--\blpage{36}
(\byear{1988})
\doiurl{10.1021/ci00057a005}
\end{barticle}
\endbibitem

%%% 40
\bibitem[\protect\citeauthoryear{Rasmussen}{2003}]{rasmussen2003gaussian}
\begin{botherref}
\oauthor{\bsnm{Rasmussen}, \binits{C.E.}}:
Gaussian {P}rocesses in machine learning,
pp. 63--71.
Springer
(2003)
\end{botherref}
\endbibitem

%%% 41
\bibitem[\protect\citeauthoryear{Kawai}{1995}]{kawai1995plastic}
\begin{barticle}
\bauthor{\bsnm{Kawai}, \binits{H.}}:
\batitle{Plastic molding materials for precision optics}.
\bjtitle{Optics}
\bvolume{24}(\bissue{2}),
\bfpage{69}--\blpage{75}
(\byear{1995})
\end{barticle}
\endbibitem

%%% 42
\bibitem[\protect\citeauthoryear{Badur et~al.}{2018}]{badur2018high}
\begin{barticle}
\bauthor{\bsnm{Badur}, \binits{T.}},
\bauthor{\bsnm{Dams}, \binits{C.}},
\bauthor{\bsnm{Hampp}, \binits{N.}}:
\batitle{High refractive index polymers by design}.
\bjtitle{Macromolecules}
\bvolume{51}(\bissue{11}),
\bfpage{4220}--\blpage{4228}
(\byear{2018})
\end{barticle}
\endbibitem

%%% 43
\bibitem[\protect\citeauthoryear{Zhang et~al.}{2023}]{zhang2023all}
\begin{barticle}
\bauthor{\bsnm{Zhang}, \binits{J.}},
\bauthor{\bsnm{Bai}, \binits{T.}},
\bauthor{\bsnm{Liu}, \binits{W.}},
\bauthor{\bsnm{Li}, \binits{M.}},
\bauthor{\bsnm{Zang}, \binits{Q.}},
\bauthor{\bsnm{Ye}, \binits{C.}},
\bauthor{\bsnm{Sun}, \binits{J.Z.}},
\bauthor{\bsnm{Shi}, \binits{Y.}},
\bauthor{\bsnm{Ling}, \binits{J.}},
\bauthor{\bsnm{Qin}, \binits{A.}},
\bauthor{\bsnm{Tang}, \binits{B.Z.}}:
\batitle{All-organic polymeric materials with high refractive index and excellent transparency}.
\bjtitle{Nature Communications}
\bvolume{14}(\bissue{1}),
\bfpage{3524}
(\byear{2023})
\end{barticle}
\endbibitem

%%% 44
\bibitem[\protect\citeauthoryear{Higashihara and Ueda}{2015}]{higashihara2015recent}
\begin{barticle}
\bauthor{\bsnm{Higashihara}, \binits{T.}},
\bauthor{\bsnm{Ueda}, \binits{M.}}:
\batitle{Recent progress in high refractive index polymers}.
\bjtitle{Macromolecules}
\bvolume{48}(\bissue{7}),
\bfpage{1915}--\blpage{1929}
(\byear{2015})
\end{barticle}
\endbibitem

%%% 45
\bibitem[\protect\citeauthoryear{Leosson and Agnarsson}{2012}]{leosson2012integrated}
\begin{barticle}
\bauthor{\bsnm{Leosson}, \binits{K.}},
\bauthor{\bsnm{Agnarsson}, \binits{B.}}:
\batitle{Integrated biophotonics with {CYTOP}}.
\bjtitle{Micromachines}
\bvolume{3}(\bissue{1}),
\bfpage{114}--\blpage{125}
(\byear{2012})
\end{barticle}
\endbibitem

%%% 46
\bibitem[\protect\citeauthoryear{Smith et~al.}{2020}]{smith2020psi4}
\begin{botherref}
\oauthor{\bsnm{Smith}, \binits{D.G.A.}},
\oauthor{\bsnm{Burns}, \binits{L.A.}},
\oauthor{\bsnm{Simmonett}, \binits{A.C.}},
\oauthor{\bsnm{Parrish}, \binits{R.M.}},
\oauthor{\bsnm{Schieber}, \binits{M.C.}},
\oauthor{\bsnm{Galvelis}, \binits{R.}},
\oauthor{\bsnm{Kraus}, \binits{P.}},
\oauthor{\bsnm{Kruse}, \binits{H.}},
\oauthor{\bsnm{Di~Remigio}, \binits{R.}},
\oauthor{\bsnm{Alenaizan}, \binits{A.}},
\oauthor{\bsnm{James}, \binits{A.M.}},
\oauthor{\bsnm{Lehtola}, \binits{S.}},
\oauthor{\bsnm{Misiewicz}, \binits{J.P.}},
\oauthor{\bsnm{Scheurer}, \binits{M.}},
\oauthor{\bsnm{Shaw}, \binits{R.A.}},
\oauthor{\bsnm{Schriber}, \binits{J.B.}},
\oauthor{\bsnm{Xie}, \binits{Y.}},
\oauthor{\bsnm{Glick}, \binits{Z.L.}},
\oauthor{\bsnm{Sirianni}, \binits{D.A.}},
\oauthor{\bsnm{O’Brien}, \binits{J.S.}},
\oauthor{\bsnm{Waldrop}, \binits{J.M.}},
\oauthor{\bsnm{Kumar}, \binits{A.}},
\oauthor{\bsnm{Hohenstein}, \binits{E.G.}},
\oauthor{\bsnm{Pritchard}, \binits{B.P.}},
\oauthor{\bsnm{Brooks}, \binits{B.R.}},
\oauthor{\bsnm{Schaefer}, \binits{H.F.}},
\oauthor{\bsnm{Sokolov}, \binits{A.Y.}},
\oauthor{\bsnm{Patkowski}, \binits{K.}},
\oauthor{\bsnm{DePrince}, \binits{A.E.}},
\oauthor{\bsnm{Bozkaya}, \binits{U.}},
\oauthor{\bsnm{King}, \binits{R.A.}},
\oauthor{\bsnm{Evangelista}, \binits{F.A.}},
\oauthor{\bsnm{Turney}, \binits{J.M.}},
\oauthor{\bsnm{Crawford}, \binits{T.D.}},
\oauthor{\bsnm{Sherrill}, \binits{C.D.}}:
{PSI}4 1.4: {O}pen-source software for high-throughput quantum chemistry.
The Journal of Chemical Physics
\textbf{152}(18)
(2020)
\end{botherref}
\endbibitem

%%% 47
\bibitem[\protect\citeauthoryear{Larsen et~al.}{2011}]{larsen2011molecular}
\begin{barticle}
\bauthor{\bsnm{Larsen}, \binits{G.S.}},
\bauthor{\bsnm{Lin}, \binits{P.}},
\bauthor{\bsnm{Hart}, \binits{K.E.}},
\bauthor{\bsnm{Colina}, \binits{C.M.}}:
\batitle{Molecular simulations of {PIM}-1-like polymers of intrinsic microporosity}.
\bjtitle{Macromolecules}
\bvolume{44}(\bissue{17}),
\bfpage{6944}--\blpage{6951}
(\byear{2011})
\end{barticle}
\endbibitem

%%% 48
\bibitem[\protect\citeauthoryear{Mardirossian and Head-Gordon}{2016}]{mardirossian2016omegab97m}
\begin{botherref}
\oauthor{\bsnm{Mardirossian}, \binits{N.}},
\oauthor{\bsnm{Head-Gordon}, \binits{M.}}:
$\omega${B97M-V}: A combinatorially optimized, range-separated hybrid, meta-{GGA} density functional with {VV}10 nonlocal correlation.
The Journal of Chemical Physics
\textbf{144}(21)
(2016)
\end{botherref}
\endbibitem

%%% 49
\bibitem[\protect\citeauthoryear{Grimme et~al.}{2011}]{grimme2011effect}
\begin{barticle}
\bauthor{\bsnm{Grimme}, \binits{S.}},
\bauthor{\bsnm{Ehrlich}, \binits{S.}},
\bauthor{\bsnm{Goerigk}, \binits{L.}}:
\batitle{Effect of the damping function in dispersion corrected density functional theory}.
\bjtitle{Journal of Computational Chemistry}
\bvolume{32}(\bissue{7}),
\bfpage{1456}--\blpage{1465}
(\byear{2011})
\end{barticle}
\endbibitem

%%% 50
\bibitem[\protect\citeauthoryear{Ditchfield et~al.}{1971}]{ditchfield1971self}
\begin{barticle}
\bauthor{\bsnm{Ditchfield}, \binits{R.}},
\bauthor{\bsnm{Hehre}, \binits{W.J.}},
\bauthor{\bsnm{Pople}, \binits{J.A.}}:
\batitle{Self-consistent molecular-orbital methods. {IX}. {A}n extended {G}aussian-type basis for molecular-orbital studies of organic molecules}.
\bjtitle{The Journal of Chemical Physics}
\bvolume{54}(\bissue{2}),
\bfpage{724}--\blpage{728}
(\byear{1971})
\end{barticle}
\endbibitem

%%% 51
\bibitem[\protect\citeauthoryear{Francl et~al.}{1982}]{francl1982self}
\begin{barticle}
\bauthor{\bsnm{Francl}, \binits{M.M.}},
\bauthor{\bsnm{Pietro}, \binits{W.J.}},
\bauthor{\bsnm{Hehre}, \binits{W.J.}},
\bauthor{\bsnm{Binkley}, \binits{J.S.}},
\bauthor{\bsnm{Gordon}, \binits{M.S.}},
\bauthor{\bsnm{DeFrees}, \binits{D.J.}},
\bauthor{\bsnm{Pople}, \binits{J.A.}}:
\batitle{Self-consistent molecular orbital methods. {XXIII}. {A} polarization-type basis set for second-row elements}.
\bjtitle{The Journal of Chemical Physics}
\bvolume{77}(\bissue{7}),
\bfpage{3654}--\blpage{3665}
(\byear{1982})
\end{barticle}
\endbibitem

%%% 52
\bibitem[\protect\citeauthoryear{Krishnan et~al.}{1980}]{krishnan1980self}
\begin{barticle}
\bauthor{\bsnm{Krishnan}, \binits{R.}},
\bauthor{\bsnm{Binkley}, \binits{J.S.}},
\bauthor{\bsnm{Seeger}, \binits{R.}},
\bauthor{\bsnm{Pople}, \binits{J.A.}}:
\batitle{Self-consistent molecular orbital methods. {XX}. {A} basis set for correlated wave functions}.
\bjtitle{The Journal of Chemical Physics}
\bvolume{72}(\bissue{1}),
\bfpage{650}--\blpage{654}
(\byear{1980})
\end{barticle}
\endbibitem

%%% 53
\bibitem[\protect\citeauthoryear{McLean and Chandler}{1980}]{mclean1980contracted}
\begin{barticle}
\bauthor{\bsnm{McLean}, \binits{A.}},
\bauthor{\bsnm{Chandler}, \binits{G.}}:
\batitle{Contracted {G}aussian basis sets for molecular calculations. {I}. {S}econd row atoms, {Z}= 11--18}.
\bjtitle{The Journal of Chemical Physics}
\bvolume{72}(\bissue{10}),
\bfpage{5639}--\blpage{5648}
(\byear{1980})
\end{barticle}
\endbibitem

%%% 54
\bibitem[\protect\citeauthoryear{Binning~Jr and Curtiss}{1990}]{binning1990compact}
\begin{barticle}
\bauthor{\bsnm{Binning~Jr}, \binits{R.}},
\bauthor{\bsnm{Curtiss}, \binits{L.}}:
\batitle{Compact contracted basis sets for third-row atoms: {G}a--{K}r}.
\bjtitle{Journal of Computational Chemistry}
\bvolume{11}(\bissue{10}),
\bfpage{1206}--\blpage{1216}
(\byear{1990})
\end{barticle}
\endbibitem

%%% 55
\bibitem[\protect\citeauthoryear{Clark et~al.}{1983}]{clark1983efficient}
\begin{barticle}
\bauthor{\bsnm{Clark}, \binits{T.}},
\bauthor{\bsnm{Chandrasekhar}, \binits{J.}},
\bauthor{\bsnm{Spitznagel}, \binits{G.W.}},
\bauthor{\bsnm{Schleyer}, \binits{P.V.R.}}:
\batitle{Efficient diffuse function-augmented basis sets for anion calculations. {III}. {T}he 3-21+{G} basis set for first-row elements, {Li--F}}.
\bjtitle{Journal of Computational Chemistry}
\bvolume{4}(\bissue{3}),
\bfpage{294}--\blpage{301}
(\byear{1983})
\end{barticle}
\endbibitem

%%% 56
\bibitem[\protect\citeauthoryear{Frisch et~al.}{1984}]{frisch1984self}
\begin{barticle}
\bauthor{\bsnm{Frisch}, \binits{M.J.}},
\bauthor{\bsnm{Pople}, \binits{J.A.}},
\bauthor{\bsnm{Binkley}, \binits{J.S.}}:
\batitle{Self-consistent molecular orbital methods 25. {S}upplementary functions for {G}aussian basis sets}.
\bjtitle{The Journal of Chemical Physics}
\bvolume{80}(\bissue{7}),
\bfpage{3265}--\blpage{3269}
(\byear{1984})
\end{barticle}
\endbibitem

%%% 57
\bibitem[\protect\citeauthoryear{Wadt and Hay}{1985}]{wadt1985ab}
\begin{barticle}
\bauthor{\bsnm{Wadt}, \binits{W.R.}},
\bauthor{\bsnm{Hay}, \binits{P.J.}}:
\batitle{Ab initio effective core potentials for molecular calculations. {P}otentials for main group elements {N}a to {B}i}.
\bjtitle{The Journal of Chemical Physics}
\bvolume{82}(\bissue{1}),
\bfpage{284}--\blpage{298}
(\byear{1985})
\end{barticle}
\endbibitem

%%% 58
\bibitem[\protect\citeauthoryear{Rice and Handy}{1991}]{rice1991calculation}
\begin{barticle}
\bauthor{\bsnm{Rice}, \binits{J.E.}},
\bauthor{\bsnm{Handy}, \binits{N.C.}}:
\batitle{The calculation of frequency-dependent polarizabilities as pseudo-energy derivatives}.
\bjtitle{The Journal of Chemical Physics}
\bvolume{94}(\bissue{7}),
\bfpage{4959}--\blpage{4971}
(\byear{1991})
\end{barticle}
\endbibitem

%%% 59
\bibitem[\protect\citeauthoryear{Yanai et~al.}{2004}]{yanai2004new}
\begin{barticle}
\bauthor{\bsnm{Yanai}, \binits{T.}},
\bauthor{\bsnm{Tew}, \binits{D.P.}},
\bauthor{\bsnm{Handy}, \binits{N.C.}}:
\batitle{A new hybrid exchange--correlation functional using the coulomb-attenuating method {(CAM-B3LYP)}}.
\bjtitle{Chemical Physics Letters}
\bvolume{393}(\bissue{1-3}),
\bfpage{51}--\blpage{57}
(\byear{2004})
\end{barticle}
\endbibitem

\end{thebibliography}
\end{document}

\title{\centering {\large Supplementary Information} \\[1ex]  SPACIER: On-Demand Polymer Design with Fully Automated All-Atom Classical Molecular Dynamics Integrated into Machine Learning Pipelines}

\author*[1]{\fnm{Shun} \sur{Nanjo}}\email{nanjos@ism.ac.jp}
\author[2]{Arifin}
\author[3]{\fnm{Hayato} \sur{Maeda}}
\author[1,4]{\fnm{Yoshihiro} \sur{Hayashi}}
\author[3]{\fnm{Kan} \sur{Hatakeyama-Sato}}
\author[1]{\fnm{Ryoji} \sur{Himeno}}
\author[3]{\fnm{Teruaki} \sur{Hayakawa}}
\author*[1,4]{\fnm{Ryo} \sur{Yoshida}}\email{yoshidar@ism.ac.jp}

\affil[1]{The Graduate University for Advanced Studies, SOKENDAI, Tachikawa, Tokyo, 190-8562, Japan}
\affil[2]{RD Technology and Digital Transformation Center, JSR Corporation, Kawasaki, 210-0821, Japan}
\affil[3]{Tokyo Institute of Technology, Meguro-ku, Tokyo 152-8550, Japan}
\affil[4]{The Institute of Statistical Mathematics, Research Organization of Information and Systems, Tachikawa, Tokyo 190-8562, Japan}

\maketitle
\setcounter{figure}{0}

\begin{figure}[h]%
\renewcommand{\thefigure}{S\arabic{figure}}
\centering
\includegraphics[width=1\textwidth, page=1]{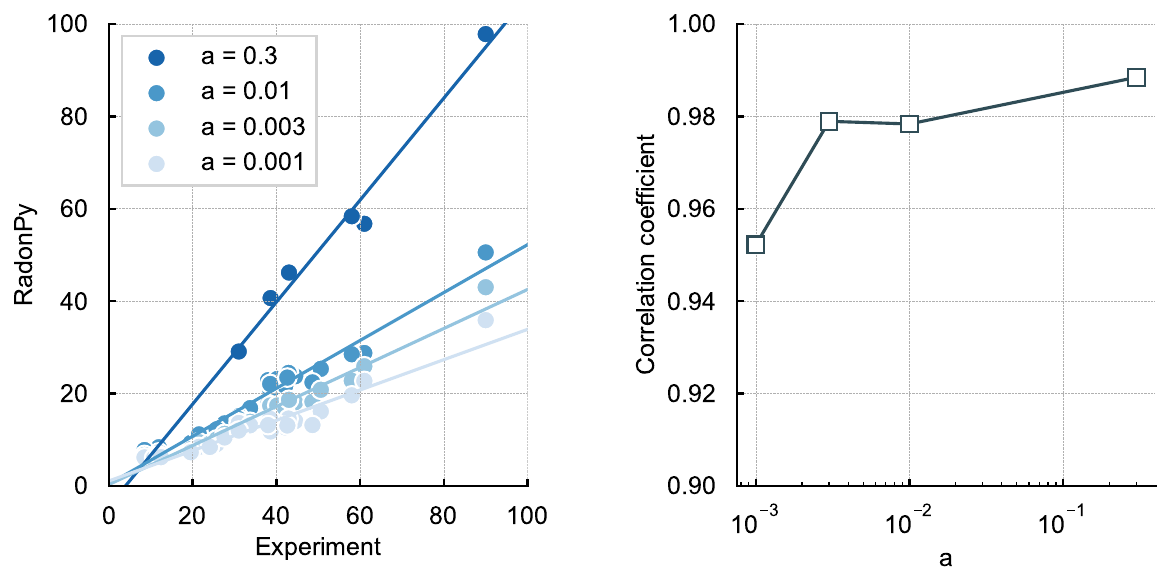}
\caption{
The dependency of the Abbe number calculated by MD simulations on the number of excited states in the TD-DFT calculations. Left: Parity plot of experimental and calculated Abbe numbers for 26 polymers, varying $a \in (0.3, 0.01, 0.003, 0.001)$, representing the proportion of excited states considered in the TD-DFT calculation relative to the total number of excited states. Right: The dependency of the correlation coefficient between experimental and calculated Abbe numbers on $a$.}
\end{figure}

\begin{figure}[h]%
\renewcommand{\thefigure}{S\arabic{figure}}
\centering
\includegraphics[width=1\textwidth, page=1]{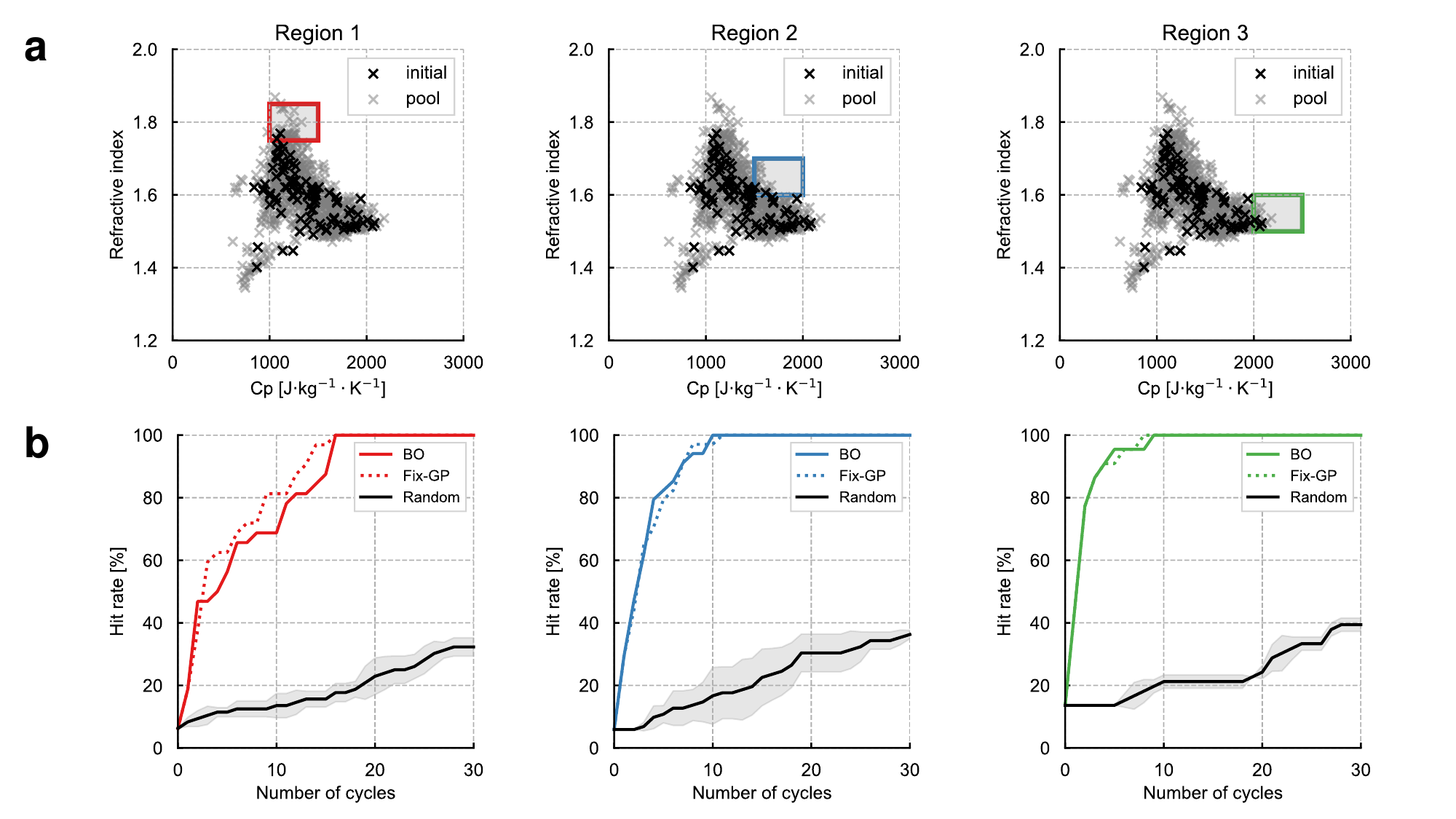}
\caption{
Results of applying SPACIER to target $C_\text{p}$ and refractive index for an initial dataset size of 100. \textbf{a} Three different target property regions (enclosed by squares) are plotted on the joint distribution of the two MD-calculated properties for all candidate polymers (gray). Initial data points are plotted in black. \textbf{b} Hit rate versus the number of BO cycles. Hit rate represents the percentage of polymers within the designated target region. ``Random'' represents the mean and standard deviation of three independent trials. }
\end{figure}

\begin{figure}[h]%
\renewcommand{\thefigure}{S\arabic{figure}}
\centering
\includegraphics[width=1\textwidth, page=2]{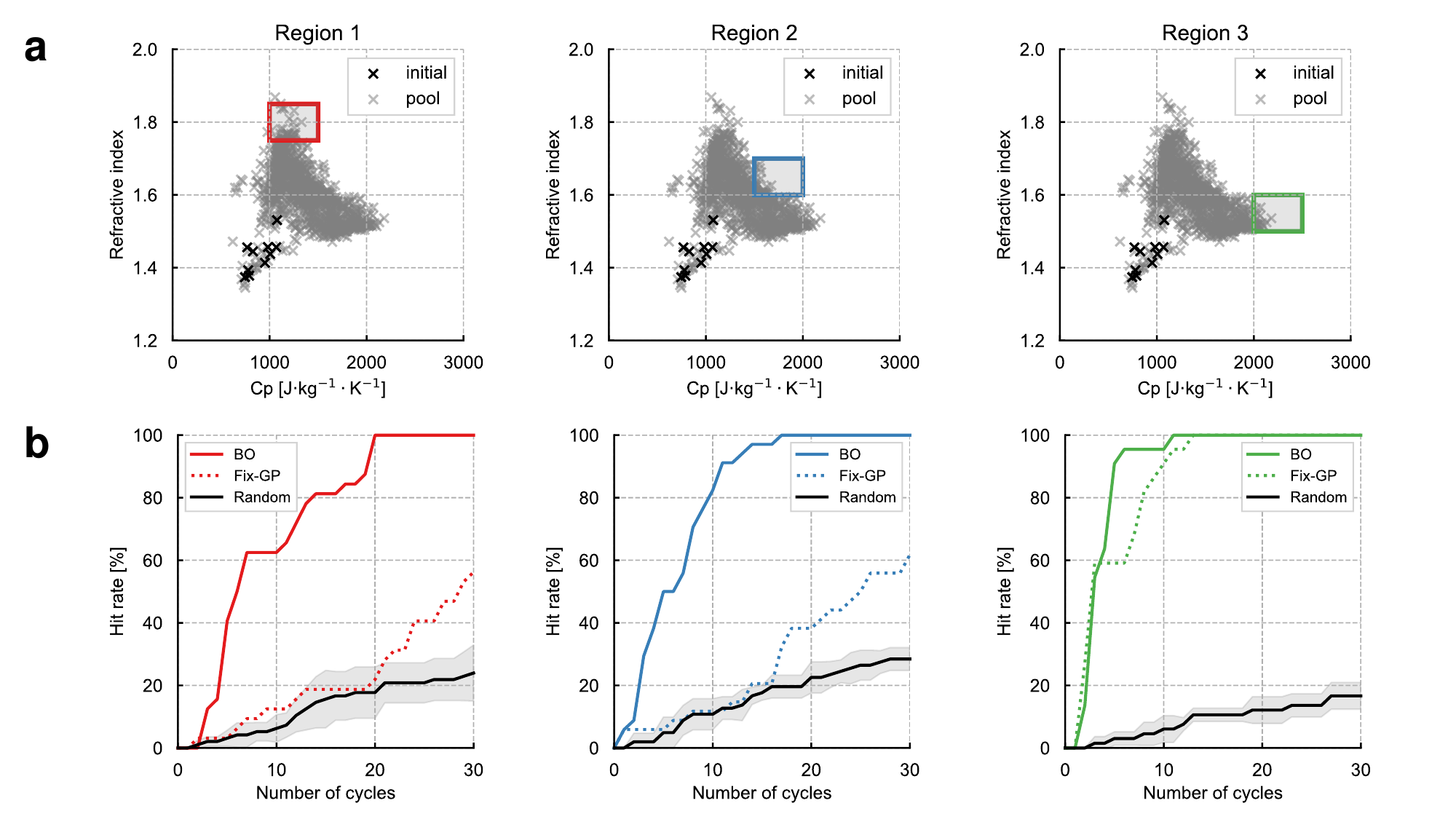}
\caption{
Results of applying SPACIER to target $C_\text{p}$ and refractive index when sampling the initial dataset from a biased region with low $C_\text{p}$ and refractive index values.}
\end{figure}

\begin{figure}[h]%
\renewcommand{\thefigure}{S\arabic{figure}}
\centering
\includegraphics[width=1\textwidth, page=3]{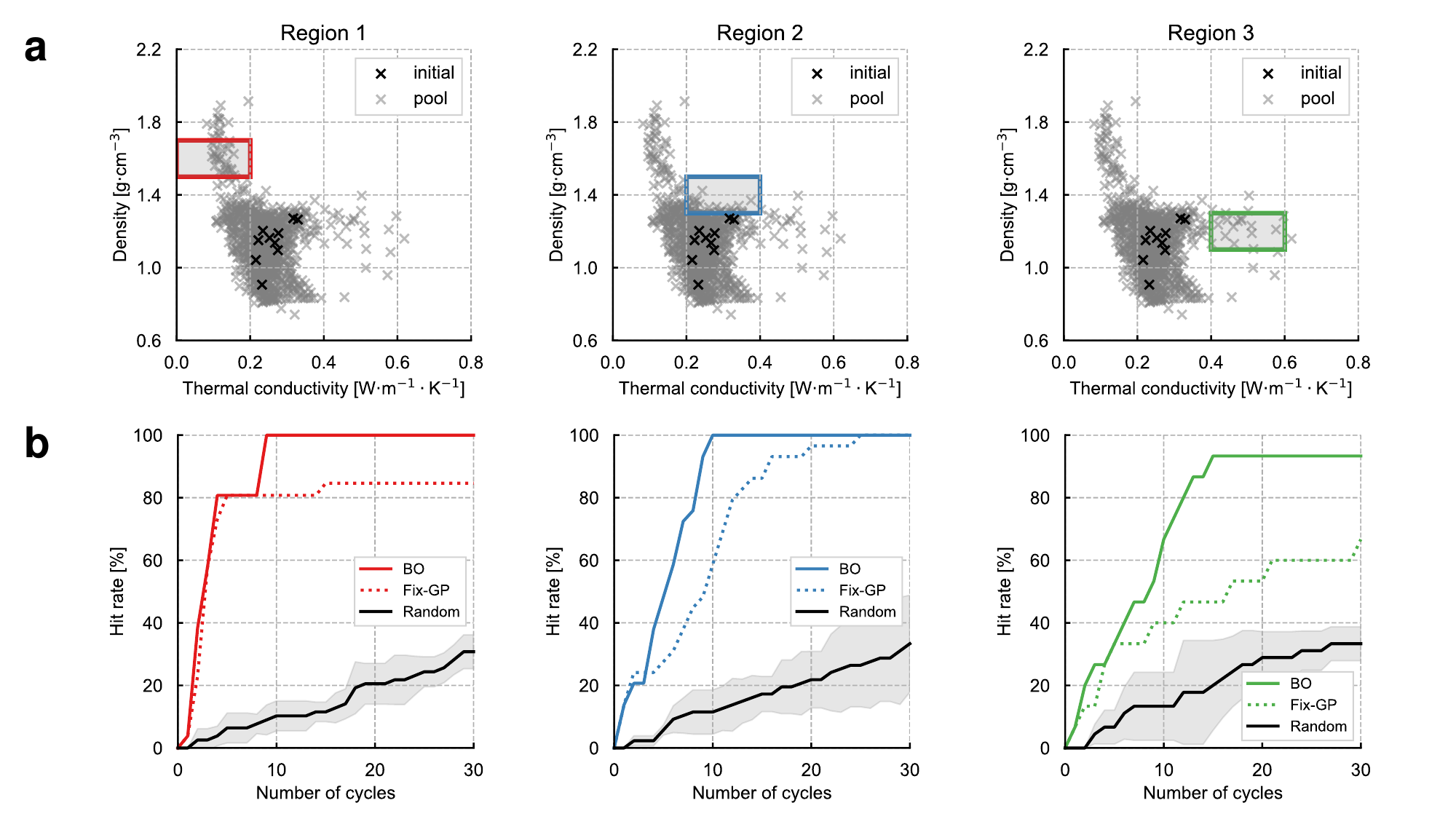}
\caption{
Results of applying SPACIER to target thermal conductivity and density. }
\end{figure}

\begin{figure}[h]%
\renewcommand{\thefigure}{S\arabic{figure}}
\centering
\includegraphics[width=1\textwidth, page=4]{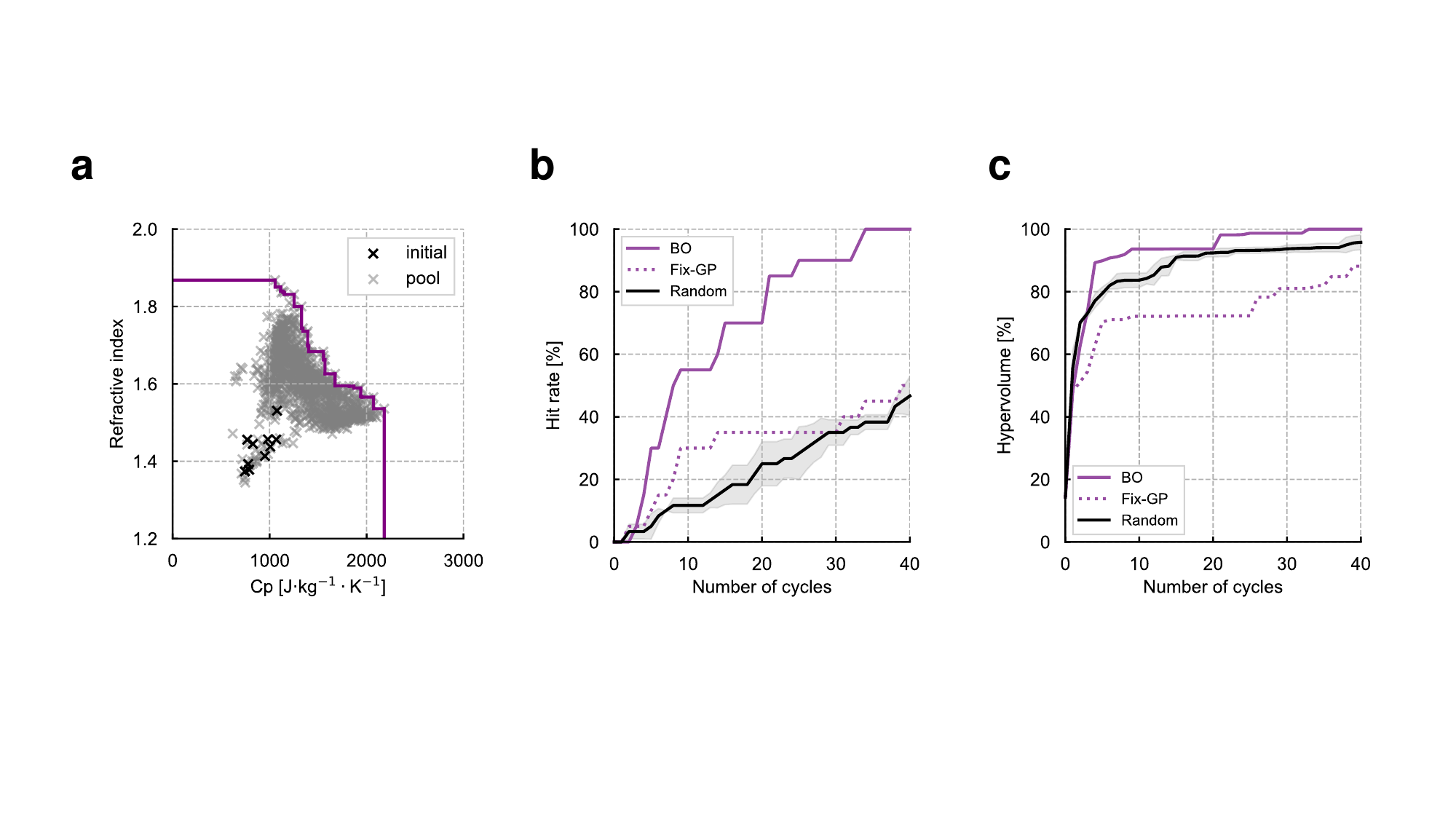}
\caption{
Results of applying EHVI to search for the optimal solution set on the Pareto boundary of $C_\text{p}$ and refractive index.
\textbf{a} Pareto boundary plotted on the joint distribution of the two MD-calculated properties for all candidate polymers (gray). Initial training data are plotted in black.
\textbf{b} Hit rate versus the number of BO cycles. Hit rate indicates the proportion of polymers falling into the optimal solution set on the Pareto boundary. ``Random'' represents the mean and standard deviation of three independent trials.
\textbf{c} Hypervolume indicator versus the number of BO cycles. Hypervolume is computed using the minimum values of the two properties as reference points.}
\end{figure}

\clearpage

\begin{figure}[h]%
\renewcommand{\thefigure}{S\arabic{figure}}
\centering
\includegraphics[width=1\textwidth, page=1]{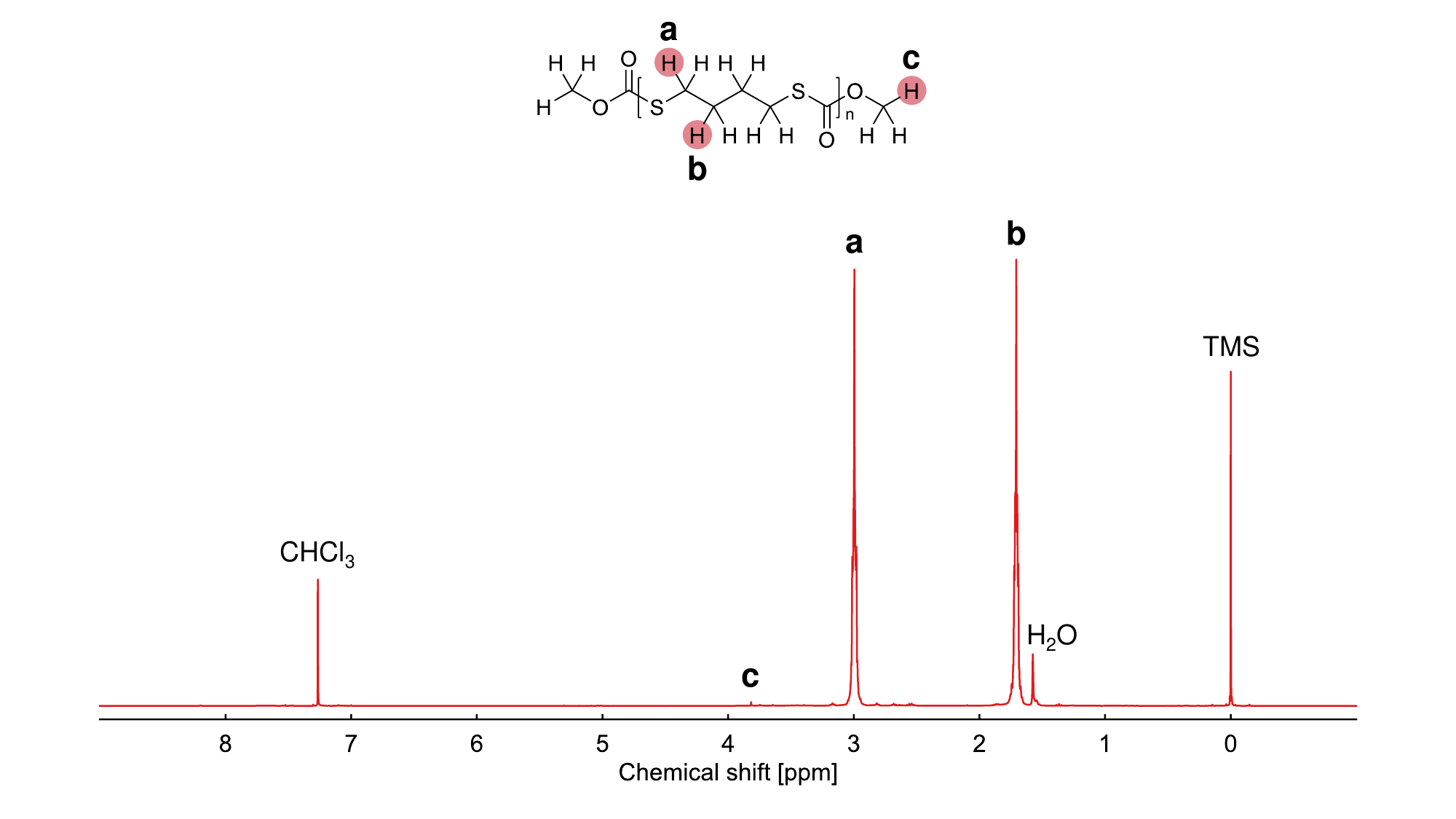}
\caption{$^{1}$H NMR spectra of \textbf{P1} in CDCl$_3$.}
\end{figure}

\begin{figure}[h]%
\renewcommand{\thefigure}{S\arabic{figure}}
\centering
\includegraphics[width=1\textwidth, page=2]{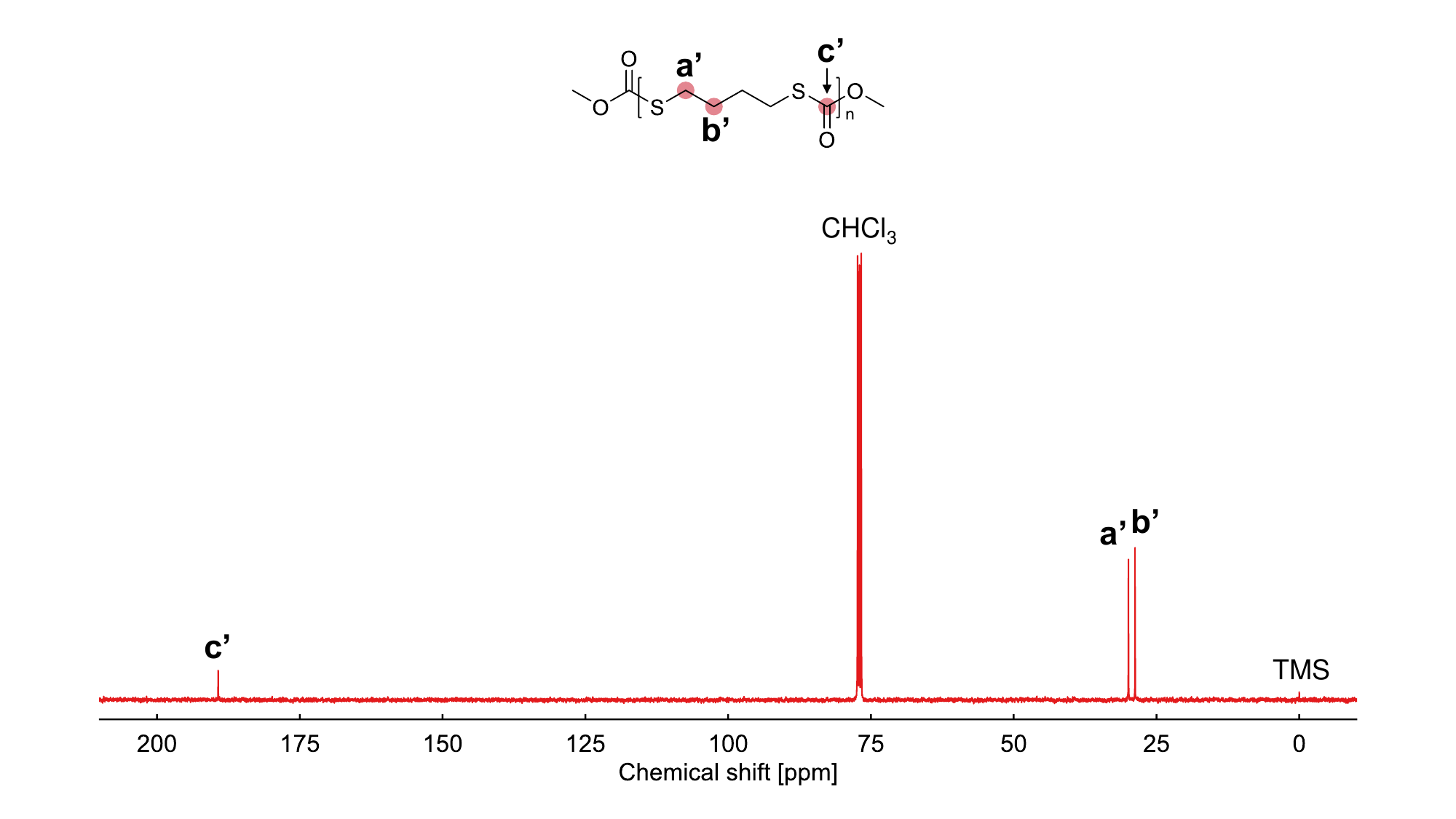}
\caption{$^{13}$C NMR spectra of \textbf{P1} in CDCl$_3$.}
\end{figure}

\begin{figure}[h]%
\renewcommand{\thefigure}{S\arabic{figure}}
\centering
\includegraphics[width=1\textwidth, page=3]{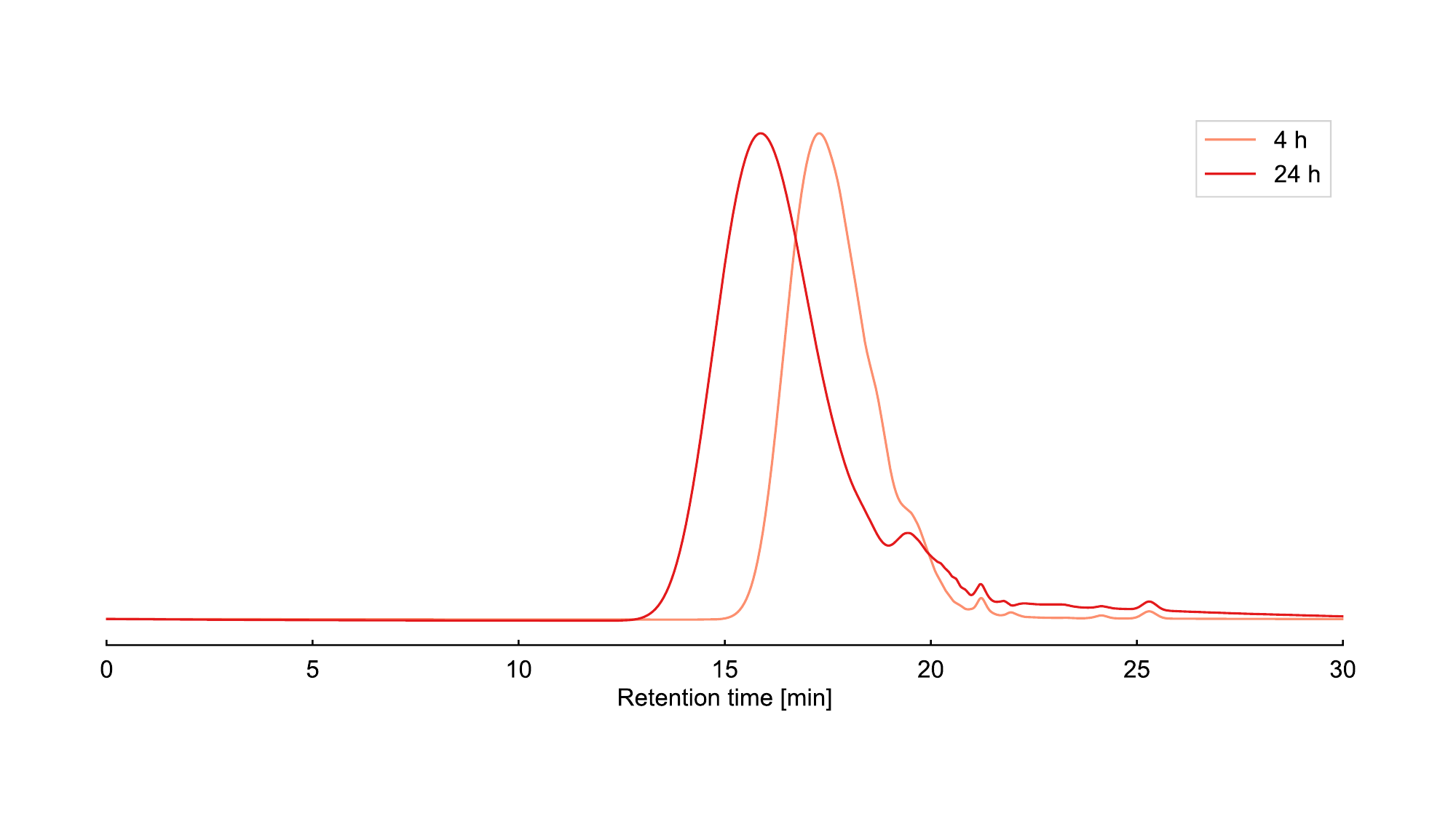}
\caption{SEC curves of \textbf{P1} after reaction times of 4 and 24 h.}
\end{figure}

\begin{figure}[h]%
\renewcommand{\thefigure}{S\arabic{figure}}
\centering
\includegraphics[width=1\textwidth, page=4]{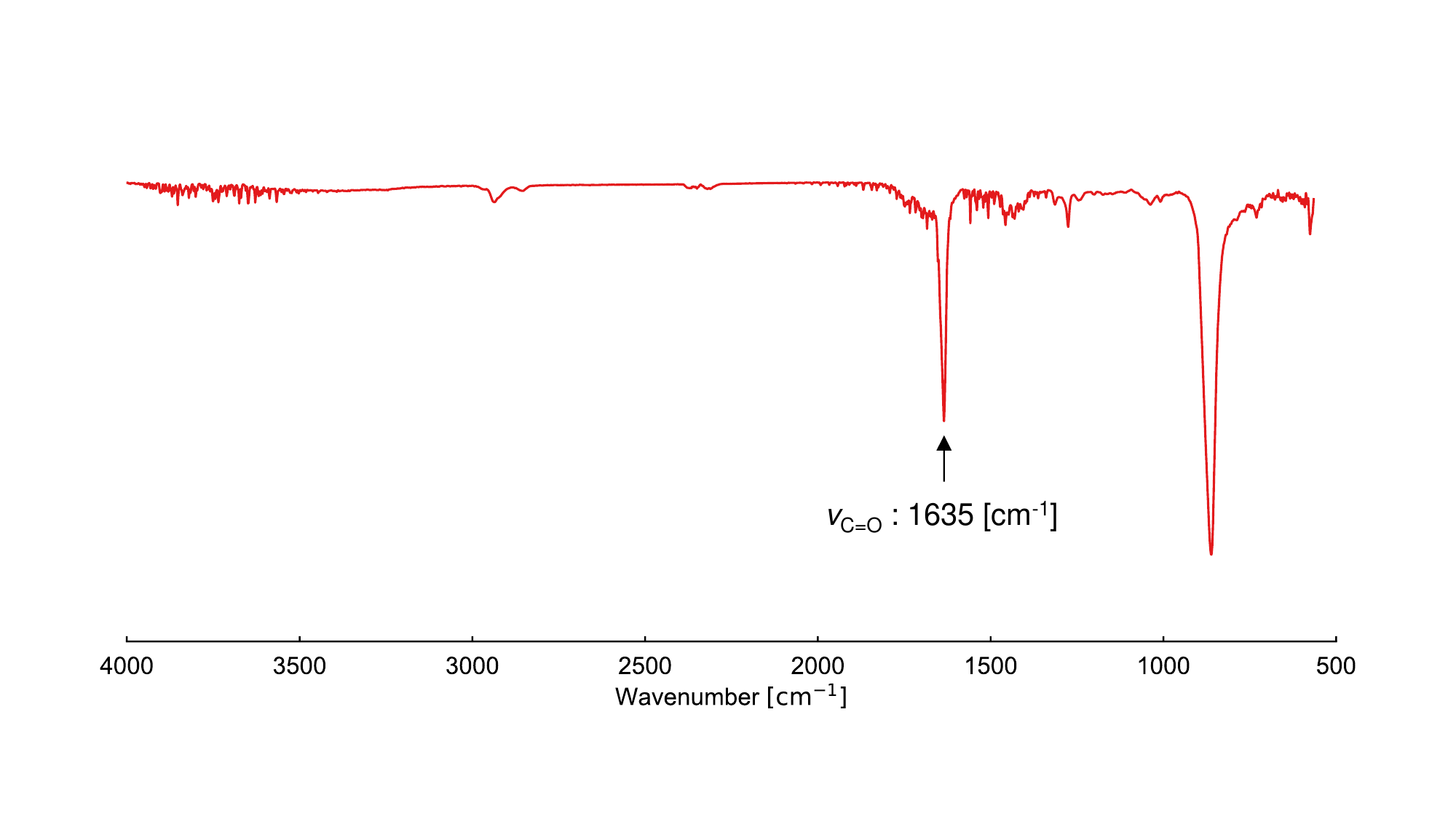}
\caption{IR spectra of \textbf{P1}.}
\end{figure}

\begin{figure}[h]%
\renewcommand{\thefigure}{S\arabic{figure}}
\centering
\includegraphics[width=1\textwidth, page=5]{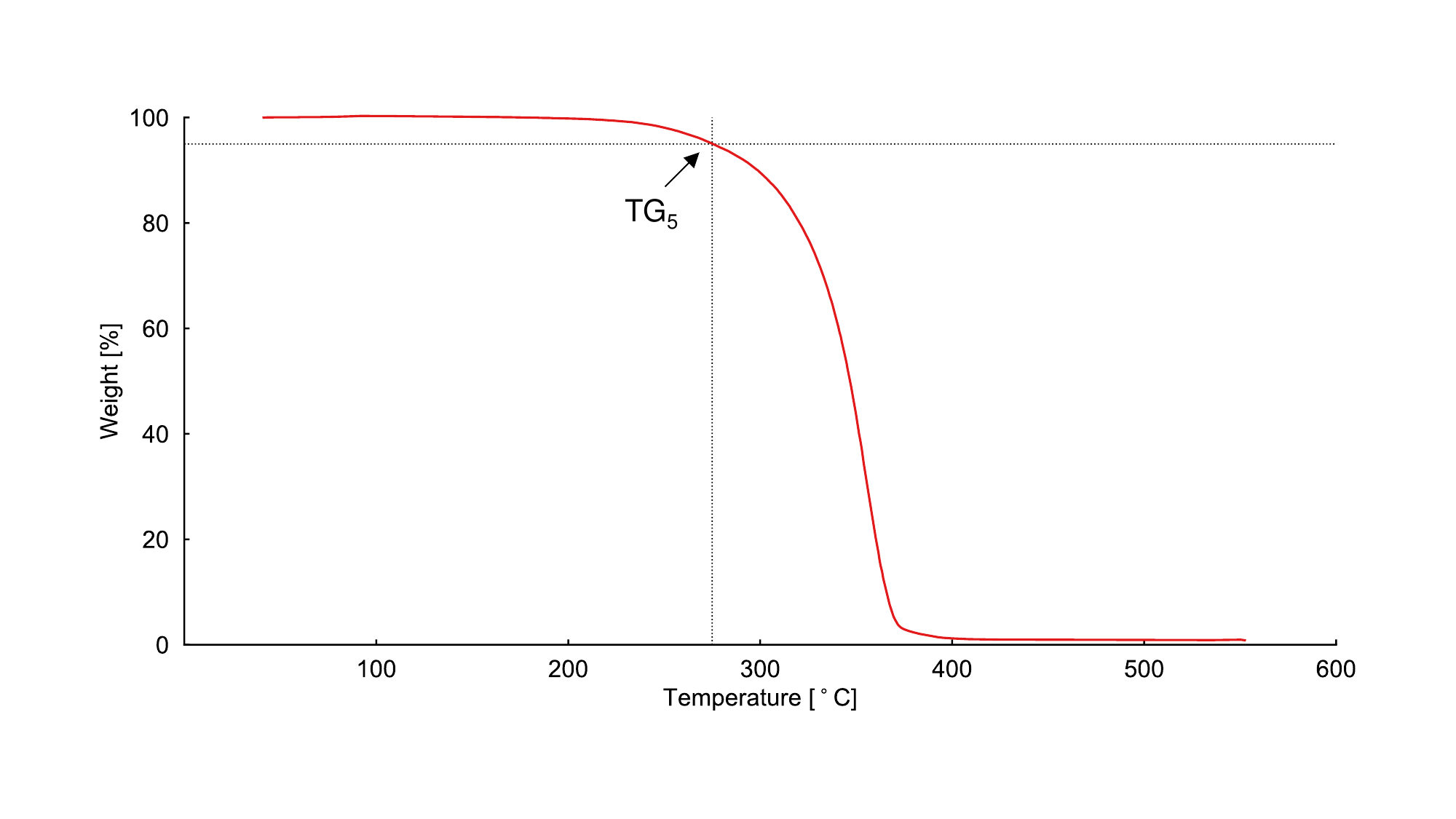}
\caption{TGA curve of \textbf{P1}.}
\end{figure}

\begin{figure}[h]%
\renewcommand{\thefigure}{S\arabic{figure}}
\centering
\includegraphics[width=1\textwidth, page=6]{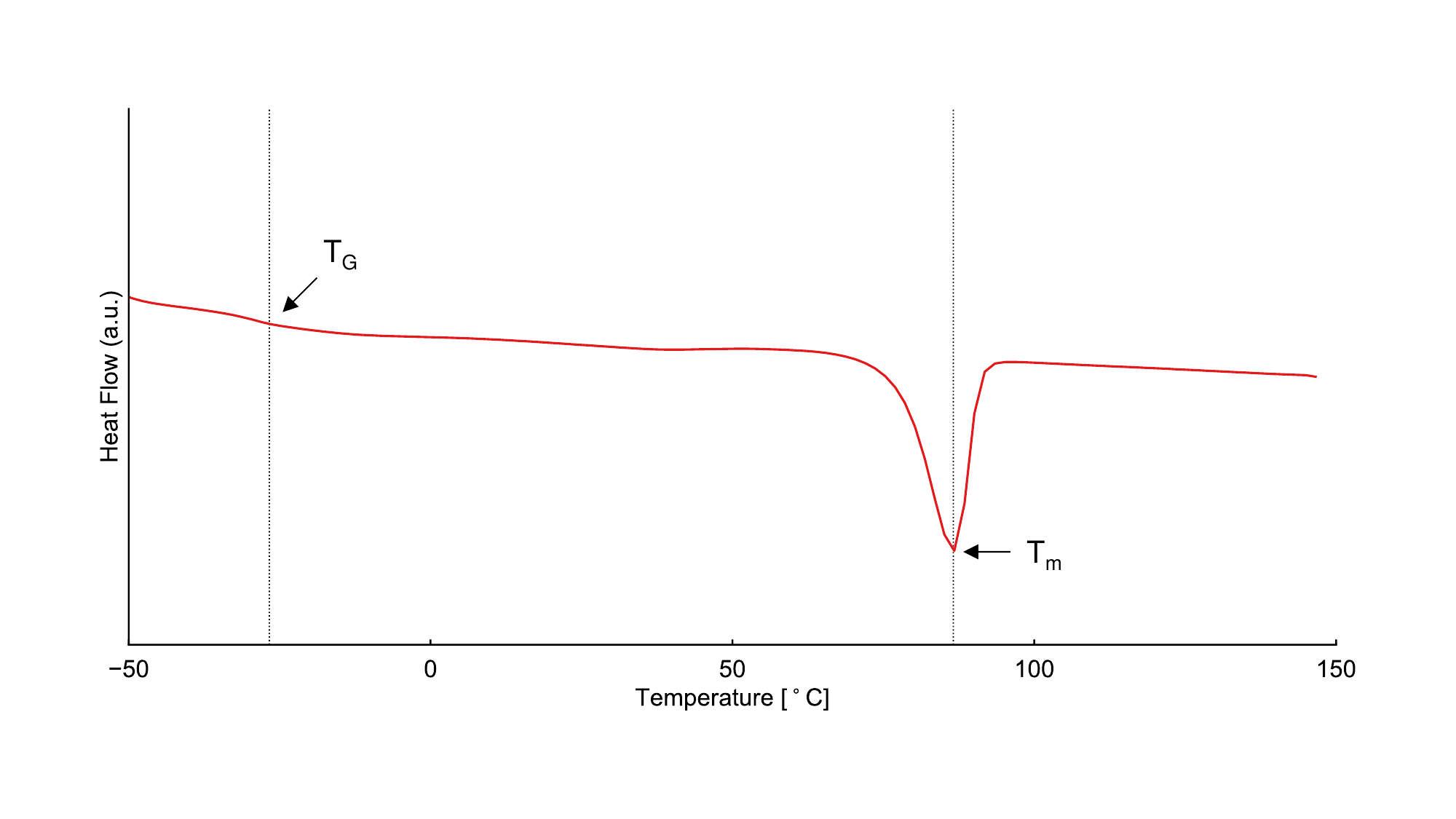}
\caption{DSC curve of \textbf{P1} during the second heating run (exo up).}
\end{figure}

\begin{figure}[h]%
\renewcommand{\thefigure}{S\arabic{figure}}
\centering
\includegraphics[width=1\textwidth, page=7]{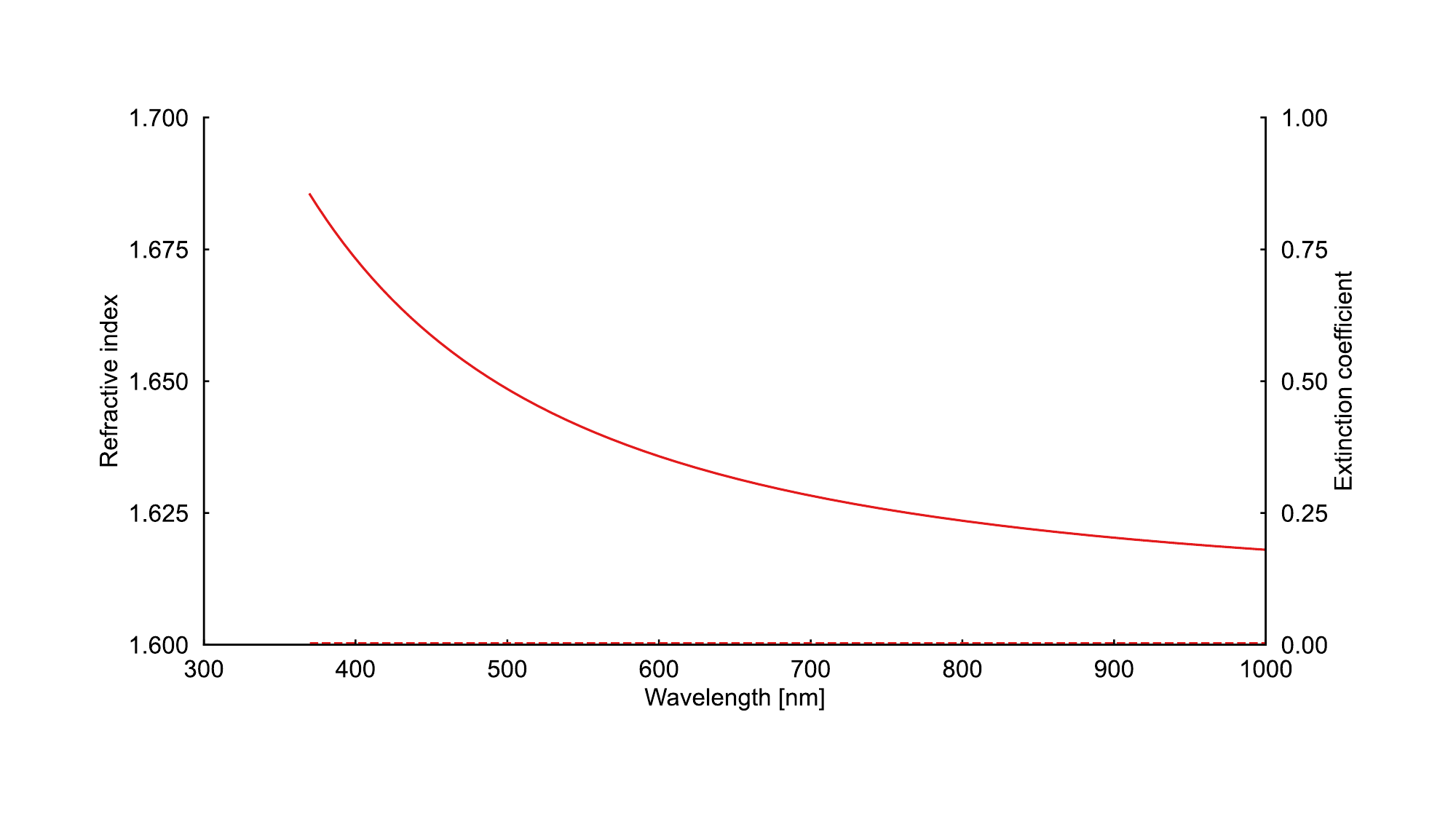}
\caption{Refractive index (solid line) and extinction coefficient (dotted line) of \textbf{P1} measured using spectroscopic ellipsometry.}
\end{figure}

\setcounter{table}{0}
\renewcommand{\thetable}{S\arabic{table}}
\begin{table}[h]
\caption{Solubility of products obtained by copolymerization of raw materials of \textbf{P2} with another monomer}
\begin{tabular*}{\textwidth}{@{\extracolsep\fill}lccc}
\toprule%
 Entry & Raw material of \textbf{P2} & Another monomer & Solubility \\ 
\hline
1 & \begin{tabular}{c} bis(2-mercaptoethyl) sulfide \\ (0.72 eq) \end{tabular} & \begin{tabular}{c} 1,6-hexanedithiol \\ (0.28 eq) \end{tabular} & + \\
2 & \begin{tabular}{c} bis(2-mercaptoethyl) sulfide \\ (0.62 eq) \end{tabular} & \begin{tabular}{c} 1,6-hexanedithiol \\ (0.38 eq) \end{tabular} & + \\ \hline
3 & \begin{tabular}{c} bis(2-mercaptoethyl) sulfide \\ (0.76 eq) \end{tabular} & \begin{tabular}{c} 3,6-dioxa-1,8-octanedithiol \\ (0.24 eq) \end{tabular}  & + \\
4 & \begin{tabular}{c} bis(2-mercaptoethyl) sulfide \\ (0.71 eq) \end{tabular} & \begin{tabular}{c} 3,6-dioxa-1,8-octanedithiol \\ (0.29 eq) \end{tabular}  & + \\ 
5 & \begin{tabular}{c} bis(2-mercaptoethyl) sulfide \\ (0.62 eq) \end{tabular} & \begin{tabular}{c} 3,6-dioxa-1,8-octanedithiol \\ (0.38 eq) \end{tabular}  & + \\
\hline
6 & \begin{tabular}{c} bis(2-mercaptoethyl) sulfide \\ (0.71 eq) \end{tabular} & \begin{tabular}{c} 1,4-cyclohexanediol \\ (0.29 eq) \end{tabular}  & + \\
7 & \begin{tabular}{c} bis(2-mercaptoethyl) sulfide \\ (0.67 eq) \end{tabular} & \begin{tabular}{c} 1,4-cyclohexanediol \\ (0.33 eq) \end{tabular}  & + \\ 
8 & \begin{tabular}{c} bis(2-mercaptoethyl) sulfide \\ (0.60 eq) \end{tabular} & \begin{tabular}{c} 1,4-cyclohexanediol \\ (0.40 eq) \end{tabular}  & + \\
\hline
9 & \begin{tabular}{c} bis(2-mercaptoethyl) sulfide \\ (0.83 eq) \end{tabular} & \begin{tabular}{c} 9,9‐bis(4‐hydroxyphenyl)‐fluorene  \\ (0.17 eq) \end{tabular} & - \\

10 & \begin{tabular}{c} bis(2-mercaptoethyl) sulfide \\ (0.71 eq) \end{tabular} & \begin{tabular}{c} 9,9‐bis(4‐hydroxyphenyl)‐fluorene  \\ (0.29 eq) \end{tabular} & - \\
\hline
11 & \begin{tabular}{c} bis(2-mercaptoethyl) sulfide \\ (0.48 eq) \end{tabular} & \begin{tabular}{c} 1,4-benzenedimethanethiol  \\ (0.52 eq) \end{tabular} & - \\ 
\hline
12 & \begin{tabular}{c} bis(2-mercaptoethyl) sulfide \\ (0.71 eq) \end{tabular} & \begin{tabular}{c} resorcinol  \\ (0.29 eq) \end{tabular} & - \\    
\bottomrule
\end{tabular*}
\footnotetext{+ and - indicate solubility and insolubility in chloroform.}
\end{table}

\begin{figure}[h]%
\renewcommand{\thefigure}{S\arabic{figure}}
\centering
\includegraphics[width=1\textwidth, page=8]{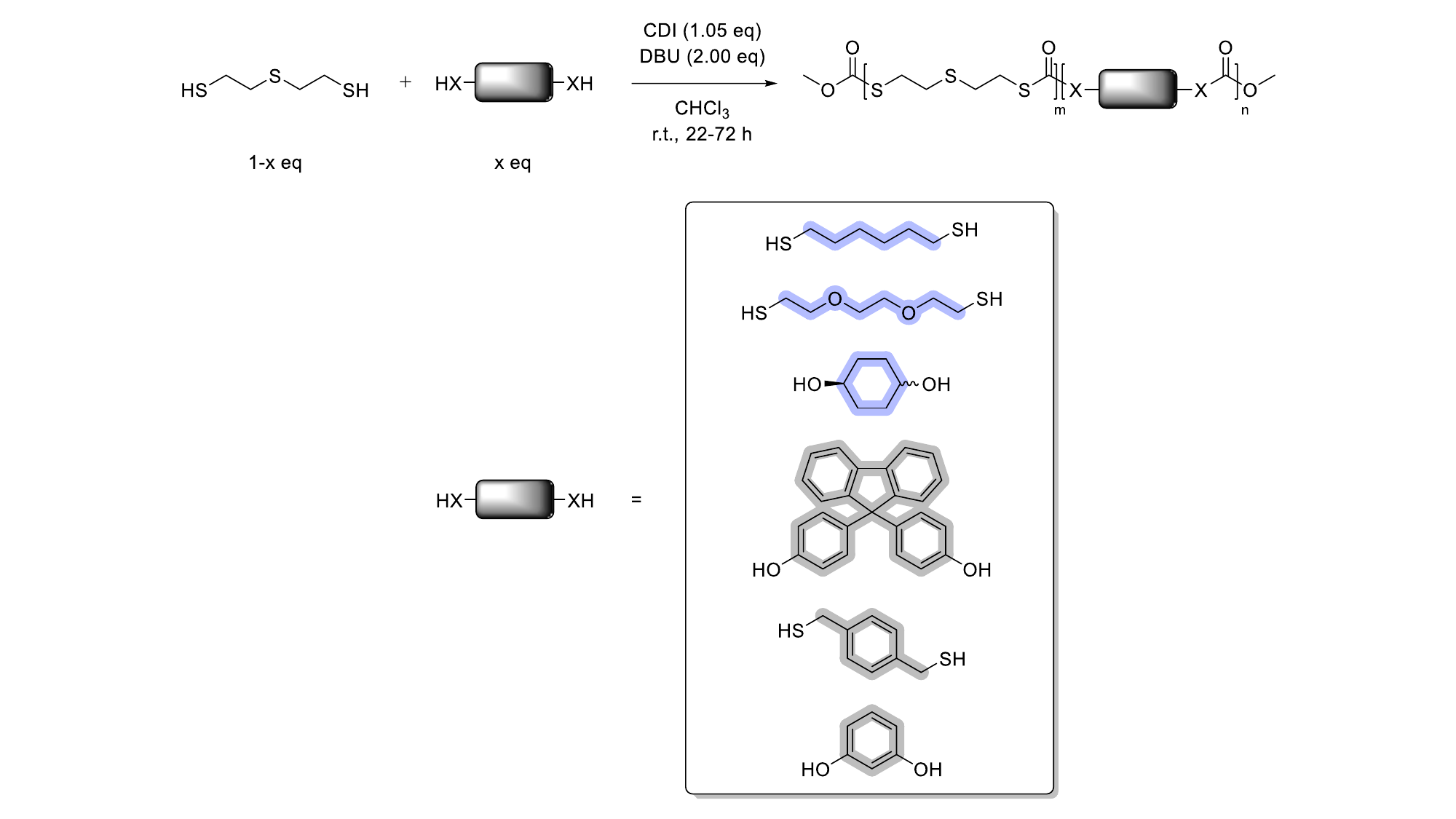}
\caption{Synthetic route for copolymers.}
\end{figure}